\documentclass[twocolumn]{aastex631} 

\received{}
\revised{}
\accepted{}

\shorttitle{Halo assembly history for SDSS galaxy groups}
\shortauthors{Lyu et al. 2024}

\graphicspath{{./}{figures/}} 

\begin{document}

\title{From Halos to Galaxies. IX. Estimate of halo assembly history for SDSS galaxy groups}


\correspondingauthor{Yingjie Peng}
\email{yjpeng@pku.edu.cn}

\author[0009-0000-7307-6362]{Cheqiu Lyu}
\affiliation{Department of Astronomy, School of Physics, Peking University, 5 Yiheyuan Road, Beijing 100871, People’s Republic of China}
\affiliation{Kavli Institute for Astronomy and Astrophysics, Peking University, 5 Yiheyuan Road, Beijing 100871, People’s Republic of China}

\author{Yingjie Peng}
\affiliation{Department of Astronomy, School of Physics, Peking University, 5 Yiheyuan Road, Beijing 100871, People’s Republic of China}
\affiliation{Kavli Institute for Astronomy and Astrophysics, Peking University, 5 Yiheyuan Road, Beijing 100871, People’s Republic of China}

\author[0000-0002-4534-3125]{Yipeng Jing}
\affiliation{Department of Astronomy, School of Physics and Astronomy, Shanghai Jiao Tong University, Shanghai 200240, People’s Republic of China}
\affiliation{Tsung-Dao Lee Institute, and Shanghai Key Laboratory for Particle Physics and Cosmology, Shanghai Jiao Tong University, Shanghai 200240, People’s Republic of China}

\author[0000-0003-3997-4606]{Xiaohu Yang}
\affiliation{Department of Astronomy, School of Physics and Astronomy, Shanghai Jiao Tong University, Shanghai 200240, People’s Republic of China}
\affiliation{Tsung-Dao Lee Institute, and Shanghai Key Laboratory for Particle Physics and Cosmology, Shanghai Jiao Tong University, Shanghai 200240, People’s Republic of China}

\author[0000-0001-6947-5846]{Luis C. Ho}
\affiliation{Kavli Institute for Astronomy and Astrophysics, Peking University, 5 Yiheyuan Road, Beijing 100871, People’s Republic of China}
\affiliation{Department of Astronomy, School of Physics, Peking University, 5 Yiheyuan Road, Beijing 100871, People’s Republic of China}

\author[0000-0002-7093-7355]{Alvio Renzini}
\affiliation{INAF--Osservatorio Astronomico di Padova, Vicolo dell’Osservatorio 5, I-35122 Padova, Italy}

\author{Dingyi Zhao}
\affiliation{Department of Astronomy, School of Physics, Peking University, 5 Yiheyuan Road, Beijing 100871, People’s Republic of China}
\affiliation{Kavli Institute for Astronomy and Astrophysics, Peking University, 5 Yiheyuan Road, Beijing 100871, People’s Republic of China}

\author[0000-0002-4803-2381]{Filippo Mannucci}
\affiliation{INAF--Osservatorio Astrofisico di Arcetri, Largo Enrico Fermi 5, I-50125 Firenze, Italy}

\author[0000-0001-5356-2419]{Houjun Mo}
\affiliation{Department of Astronomy, University of Massachusetts, Amherst MA 01003, USA}

\author[0000-0002-3775-0484]{Kai Wang}
\affiliation{Kavli Institute for Astronomy and Astrophysics, Peking University, 5 Yiheyuan Road, Beijing 100871, People’s Republic of China}

\author[0000-0002-6137-6007]{Bitao Wang}
\affiliation{Kavli Institute for Astronomy and Astrophysics, Peking University, 5 Yiheyuan Road, Beijing 100871, People’s Republic of China}

\author{Bingxiao Xu}
\affiliation{Kavli Institute for Astronomy and Astrophysics, Peking University, 5 Yiheyuan Road, Beijing 100871, People’s Republic of China}

\author[0000-0002-6961-6378]{Jing Dou}
\affiliation{School of Astronomy and Space Science, Nanjing University, Nanjing 210093, People’s Republic of China}

\author[0000-0002-9656-1800]{Anna R. Gallazzi}
\affiliation{INAF--Osservatorio Astrofisico di Arcetri, Largo Enrico Fermi 5, I-50125 Firenze, Italy}

\author[0000-0002-3890-3729]{Qiusheng Gu}
\affiliation{School of Astronomy and Space Science, Nanjing University, Nanjing 210093, People’s Republic of China}

\author[0000-0002-4985-3819]{Roberto Maiolino}
\affiliation{Cavendish Laboratory, University of Cambridge, 19 J.J. Thomson Avenue, Cambridge, CB3 0HE, UK}
\affiliation{Kavli Institute for Cosmology, University of Cambridge, Madingley Road, Cambridge, CB3 0HA, UK}
\affiliation{Department of Physics and Astronomy, University College London, Gower Street, London WC1E 6BT, UK}

\author[0000-0003-1588-9394]{Enci Wang}
\affiliation{CAS Key Laboratory for Research in Galaxies and Cosmology, Department of Astronomy, University of Science and Technology of China, Hefei, Anhui 230026, China}
\affiliation{School of Astronomy and Space Science, University of Science and Technology of China, Hefei 230026, China}

\author[0000-0003-3564-6437]{Feng Yuan}
\affiliation{Center for Astronomy and Astrophysics and Department of Physics, Fudan University, Shanghai 200438, People’s Republic of China}
\affiliation{Key Laboratory for Research in Galaxies and Cosmology, Shanghai Astronomical Observatory, Chinese Academy of Sciences, 80 Nandan Road, Shanghai 200030, People’s Republic of China}
\affiliation{University of Chinese Academy of Sciences, No. 19A Yuquan Road, Beijing 100049, People’s Republic of China}

\begin{abstract}

The properties of the galaxies are tightly connected to their host halo mass and halo assembly history. Accurate measurement of the halo assembly history in observation is challenging but crucial to the understanding of galaxy formation and evolution. The stellar-to-halo mass ratio ($M_*/M_{\mathrm{h}}$) for the centrals has often been used to indicate the halo assembly time $t_{\mathrm{h,50}}$ of the group, where $t_{\mathrm{h,50}}$ is the lookback time at which a halo has assembled half of its present-day virial mass. Using mock data from the semi-analytic models, we find that $M_*/M_{\mathrm{h}}$ shows a significant scatter with $t_{\mathrm{h,50}}$, with a strong systematic difference between the group with a star-forming central (blue group) and passive central (red group). To improve the accuracy, we develop machine-learning models to estimate $t_{\mathrm{h,50}}$ for galaxy groups using only observable quantities in the mocks. Since star-formation quenching will decouple the co-growth of the dark matter and baryon, we train our models separately for blue and red groups. Our models have successfully recovered $t_{\mathrm{h,50}}$, within an accuracy of $\sim$ 1.09 Gyr. With careful calibrations of individual observable quantities in the mocks with SDSS observations, we apply the trained models to the SDSS Yang et al. groups and derive the $t_{\mathrm{h,50}}$ for each group for the first time. The derived SDSS $t_{\mathrm{h,50}}$ distributions are in good agreement with that in the mocks, in particular for blue groups. The derived halo assembly history, together with the halo mass, make an important step forward in studying the halo-galaxy connections in observation.

\end{abstract}


~\\

\section{Introduction}

In the $\Lambda$CDM cosmological framework, galaxies are thought to form and reside within dark matter halos, intricately linking the characteristics of galaxy groups to the properties of their host halos. The halo-galaxy connections are profoundly significant, offering invaluable insights into the large-scale structure of the Universe and the intricate processes that drive galaxy formation and evolution \citep[see][for a comprehensive review]{Wechsler2018}. It is widely established that key halo properties, such as virial radius, assembly history, clustering patterns, and the formation of large-scale structures, wield considerable influence over the present-day features of galaxies. Particularly, the halo mass plays a crucial role in regulating various physical processes that shape the growth and evolution of galaxies, including star formation and quenching \citep[e.g.,][]{Gao2005, Gao2007, Diemer2017, Tinker2018, Wechsler2018, Matthee2019}.

Beyond halo mass, the secondary effects in the halo-galaxy connections have been extensively explored through hydrodynamical simulations and semi-analytic models \citep[e.g.,][]{Yang2006, Yang2012, Hearin2013, Hearin2014, Watson2015, Moster2018, Behroozi2019}. Among the various halo properties, the assembly history of the halo emerges as a pivotal determinant of galaxy evolution. Numerous studies have revealed that, at a given halo mass, there are significant differences in galaxy properties, such as stellar mass, gas content, black hole properties, color distribution, specific star formation rate, and quenched fractions, between central galaxies inhabiting early-formed halos and those in late-formed halos \citep[e.g.,][]{Zehavi2018, Matthee2017, Artale2018, Matthee2019, Davies2020, Davies2021, Cui2021, Lyu2023}. However, accurately describing and quantifying the halo assembly history presents challenging tasks.

In the realm of simulations, the assembly history of dark matter halos is commonly characterized by a singular parameter: halo assembly time. This parameter, typically defined as the moment when a halo reaches half of its final virial mass, provides a crucial framework for exploring halo evolution \citep[e.g.,][]{Lacey1993,Lemson1999,vandenBosch2002,Gao2005}. Moreover, several studies have incorporated various formation times to capture the intricate histories of individual halos \citep[e.g.,][]{Gao2005,Wechsler2006,Li2008}. Leveraging these assembly times in hydrodynamical simulations and semi-analytic models proves invaluable, as they allow for the integration of unobservable halo properties, facilitating the successful modeling of galaxy properties consistent with observation data \citep[e.g.,][]{Henriques2015,Pillepich2018,Dave2019}. A primary strength of hydrodynamical simulations and semi-analytic models lies in their ability to track galaxy evolution and compare it with the evolution of their dark matter halos and the broader large-scale environment \citep[e.g.,][]{Matthee2019}. However, given that many halo properties associated with assembly time elude direct observation with current techniques, assessing their impact on galaxy evolution purely through observation data presents a formidable challenge. Thus arises the question: How can we accurately infer halo assembly time from observable properties?

On the observation front, some studies have attempted to estimate halo assembly time using observable proxies. These proxies, such as the stellar-to-halo mass ratio and features of the halo's large-scale environment, are derived from scaling relations grounded in theoretical frameworks or simulations \citep[e.g.,][]{Lim2016, Tojeiro2017, Bradshaw2020, Wang2023}. However, beyond the dominant influence of halo mass, the correlation between a single galaxy group property and halo assembly time represents a secondary effect prone to significant scatter due to various physical processes. This variability may result in inaccurate estimations of halo assembly time. Furthermore, the complete assembly history of halos remains beyond current observational capabilities, and identifying a model-independent, single observable proxy presents a non-trivial challenge. Hence, fully leveraging observable galaxy and group properties is imperative for achieving a more accurate estimation of halo assembly time.

In physics, the growth of star-forming central galaxies and their host halos unfolds continuously over time. While passive central galaxies ceased star formation at a specific redshift, maintaining a relatively constant stellar mass unless further mergers occur. Meanwhile, their halos continue to grow by merging with smaller halos, regardless of the central galaxies' star formation status. This straightforward scenario, proposed by \cite{Peng2012} and further depicted in Figure 2 of \cite{Man2019}, sheds light on the mass relationships between central galaxies and their halos. In semi-analytic models, for instance, \citet{Lyu2023} demonstrated that the halo assembly history significantly influences the stellar mass, star formation status, and star formation history (SFH) of the central galaxy at a given present-day halo mass. Consequently, galaxy properties such as star formation rate (SFR), color, and stellar age can be traced back to the underlying properties of their host halos through the halo-galaxy connections. Recently, \cite{Zhao2024} trained machine-learning models using various galaxy and group properties, including information on star formation and quenching history, to estimate the dark matter halo mass of galaxy groups in mock data. This innovative approach led to a notable improvement in the accuracy of halo mass measurements. By applying these models to observation data in Sloan Digital Sky Survey (SDSS) DR7 \citep{Abazajian2009}, they obtained accurate measurements of the dark matter halo mass for SDSS groups, with resulting halo mass functions in good agreement with simulations and theoretical predictions. Beyond halo mass, the assembly time of the halo also plays a pivotal role in determining galaxy evolution. Therefore, it is worthwhile to explore the use of observable properties of central galaxies, alongside halo mass, to quantify whether their host halos assembled earlier or later in cosmic history.

Considering the importance of halo assembly time in governing galaxy evolution, it is imperative to explore the feasibility of inferring this parameter from observable properties of central galaxies. Although direct observation of assembly times is challenging due to technical limitations, the distinct properties of local central galaxies in halos with varying assembly times offer a promising avenue for developing a proxy to trace the assembly history of their host halos. In this work, we aim to demonstrate the feasibility of estimating halo assembly time by using the halo mass and the properties of local central galaxies. We describe our data and sample selection in Section \ref{data}, followed by data pre-processing and machine-learning methods in Section \ref{method}. Subsequently, we present the results obtained from our models in Section \ref{result} and discuss their feasibility and limitations in Section \ref{discussion}. Finally, we summarize our work in Section \ref{conclusion}. Throughout this work, we adopt $t_{\mathrm{h,50}}$ as the proxy for halo assembly time, defined as the lookback time at which a halo has assembled half of its present-day virial mass. We employ \citet{Kroupa2001} initial mass function (IMF) and adopt the following cosmological parameters: $\Omega_{\mathrm{m}}= 0.3$, $\Omega_{\mathrm{\Lambda}} = 0.7$, and $H_0=70$ km s$^{-1}$ Mpc$^{-1}$.

\section{Data}\label{data}

\subsection{L-GALAXIES}\label{LG}

L-GALAXIES is a semi-analytic galaxy formation model that leverages a suite of empirical equations to simulate the formation and evolution of galaxies based on dark matter halo merger trees \citep{Guo2011, Henriques2015}. It captures the complex baryonic cycle during galaxy formation and evolution, including various physical processes such as star formation, gas cooling, supernova feedback, and black hole growth. These physical processes are parameterized by a set of recipes, and the free parameters are carefully calibrated to reproduce key observational statistics, such as stellar mass function and quenched fraction. 

In this work, we use the version of \citet[][hereafter H15]{Henriques2015}, which is implemented on the subhalo merger trees obtained from the Millennium simulation \citep{Springel2005} scaled to first-year Planck cosmology \citep{PlanckCollaboration2014}. The simulation box has a side length of $500h^{-1}\mathrm{Mpc}$, and each dark matter particle has a mass of $8.6\times10^{8}h^{-1}\mathrm{M_\odot}$. Dark matter halos are identified using the friends-of-friends (FoF) method \citep{Davis1985}, and subhalos are identified with the SUBFIND algorithm \citep{Springel2001}. The central galaxy is attributed to the most massive subhalo in each FoF halo, which is designated as the main halo, while the remaining subhalos/galaxies are labeled as satellite ones. As the subhalo merger trees are constructed by identifying the unique descendant for each subhalo \citep{Springel2005}, we can trace the main branch by recursively identifying the most massive progenitor subhalo to build the halo assembly history. In this work, the value of halo assembly time $t_{\mathrm{h,50}}$ is calculated by linear interpolation between two lookback times that include the time at which half of the present-day virial mass (defined by $m_\mathrm{crit200}$ of the FOF group) was assembled.

Compared to hydrodynamical simulations, semi-analytic models are computationally efficient and offer better statistics due to their larger box volume. Furthermore, the stellar mass function of L-GALAXIES exhibits a good agreement with observations up to $z \sim 2$. Additionally, L-GALAXIES yields well-matched distributions of color, specific star formation rate, and luminosity-weighted stellar ages across the stellar mass range of $8.0\leq\log M_{*} / M_{\odot}\leq12.0$ \citep{Henriques2015}. Due to the resolution limit of the dark matter particles in the Millennium simulations, we select central galaxies with stellar mass $\log (M_{*}/\mathrm{M_\odot}) > 9.5$, as recommended by H15, in the snapshots within 0.01<$z$<0.20 in L-GALAXIES as mock data. We convert the \citet{Chabrier2003} IMF used in L-GALAXIES to \citet{Kroupa2001} IMF.

\subsection{Galaxy and group properties} \label{Ggp}

In this work, central galaxy properties that potentially trace back to halo assembly history can be used to estimate the halo assembly time of each galaxy group based on observations. Similar to \citet{Zhao2024}, we establish selection criteria for galaxy and group properties: (1) availability in both observations and simulations/semi-analytic models; (2) consistency between observations and simulations/semi-analytic models; (3) relevance to the assembly history of central galaxies and their host halos. This approach yields 16 qualifying quantities, summarized in Table \ref{parameters}. Among them, the stellar mass of the central galaxy ($M_*$) and the total stellar mass in the galaxy group ($M_{\mathrm{tot}}$) are commonly employed in standard abundance matching techniques \citep[e.g.,][]{Yang2005, Yang2007}. Quantities such as SFR and color indices offer insights into the SFH of central galaxies, reflecting their halo assembly history. Color indices, readily accessible in contemporary sky surveys, provide significant clues regarding the historical contribution to cosmic star formation \citep[e.g.,][]{Peterken2021}. Moreover, the stellar age of the central galaxy serves as a proxy of the quenching epoch of the red centrals, further linking to halo assembly history \citep[e.g.,][]{Lacerna2011}.

\begin{deluxetable*}{lcccc}
	\tablenum{1}
	\tablecaption{Galaxy and group properties both for L-GALAXIES and observation as model input quantities. An example of the mock data with assigned $1\sigma$ observational uncertainties (measurement noise) within 0.01<$z$<0.04 is shown here. Observational uncertainties at larger redshifts can be found in a detailed table in \citet{Zhao2024}. \label{parameters}}
	\tablewidth{0pt}
	\tablehead{\colhead{Quantity} & \colhead{Unit} & \colhead{Noise (blue, red)} & \colhead{Notes} }
	\startdata
	$\log(M_{\mathrm{*}})$& $\log(\mathrm{M_{\odot}})$&0.1,0.1 dex& stellar mass of the central galaxy (the most massive galaxy in its host halo)\\
	$\log(M_\mathrm{h})$ & $\log(\mathrm{M_{\odot}})$&0.116,0.217 dex& halo mass (the virial mass defined by $m_\mathrm{crit200}$ of the FOF group)\\	
	$\log(M_{\mathrm{tot}})$ & $\log(\mathrm{M_{\odot}})$&from error propagation& total stellar mass of the galaxies above mass threshold in a galaxy group\\
	$\log(\mathrm{SFR})$ & $\log(\mathrm{M_\odot}/\mathrm{yr})$&0.3,0.7 dex& star formation rate of the central galaxy \\
	age & Gyr &1.1,1.9& $r$-band luminosity-weighted stellar age of the central galaxy \\
	$u-g, u-r,$ & - &see \citet{Zhao2024}& ten color indices of SDSS $u, g, r, i, z$ band rest-frame absolute magnitude \\$..., i-z$&&& with dust extinction of the central galaxy
	\\
	$\log(M_*/M_{\mathrm{h}})$ & - &from error propagation& stellar-to-halo mass ratio of the central galaxy\\
	\enddata
\end{deluxetable*}

We utilize the group catalog of \citet{Yang2007}, which is based on the model magnitude and constructed from the NYU-VAGC DR7 \citep{Blanton2005} using an adaptive halo-based group finder. This group finder method starts with an assumed mass-to-light ratio to assign a tentative mass to each group. This mass is used to estimate the size and velocity dispersion of the underlying halo and determine group membership (in redshift space). This procedure is repeated until no further changes occur in group memberships \citep{Yang2005,Yang2006}. The stellar mass and SFR are adopted from the publicly available MPA-JHU DR7 catalog and are converted to a \citet{Chabrier2003} IMF. Stellar mass is obtained through Bayesian methodology by fitting the photometry \citep{Kauffmann2003a}, while the SFR is estimated from H$\alpha$ emission \citep{Brinchmann2004} and aperture-corrected using photometry \citep{Salim2007}. The total stellar mass is the sum of all the stellar mass of the galaxies that are larger than the stellar mass threshold (defined in Figure \ref{MassRedshift} of below Section \ref{DataPreProcessing}) in a group. The $r$-band luminosity-weighted stellar age is adopted from \citet{Gallazzi2005, Gallazzi2021} by fitting spectral absorption features. The absolute magnitudes ($u,g,r,i,z$) of galaxies are obtained from the NYU-VAGC DR7 {\ttfamily k-corrections} table and re-scaled to the cosmological model used in this work. For the measurement errors of these quantities above, we adopt the values derived in \citet{Zhao2024}. In addition, we use the halo mass catalog from \citet{Zhao2024}, who developed machine-learning regression models based on mock data and applied them to the SDSS DR7 galaxy group sample. The estimation errors of the halo mass are the standard deviation of residuals in test sets of various machine-learning models in \citet{Zhao2024}. For instance, we use 0.116 dex and 0.217 dex as errors of the halo mass for blue and red group samples within 0.01<$z$<0.04, respectively.

\section{Method}\label{method}

\begin{figure*}[htb!] 
	\centering
	\includegraphics[width=0.48\textwidth]{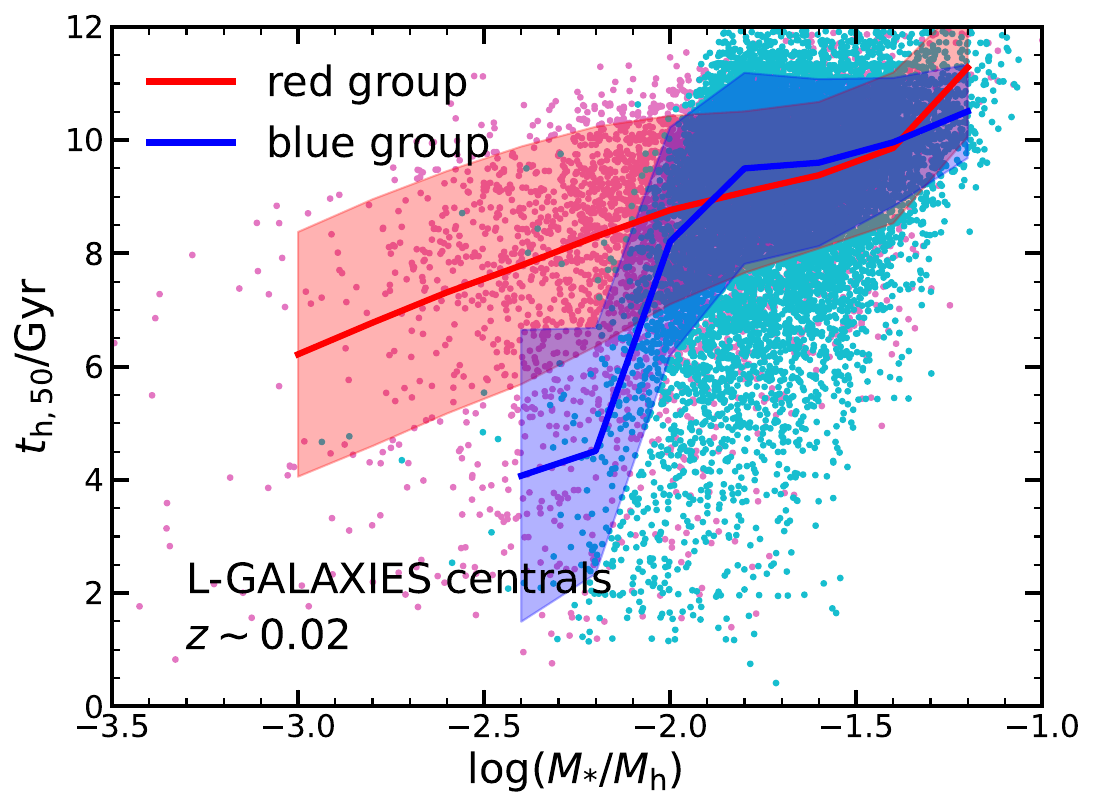}
	\includegraphics[width=0.48\textwidth]{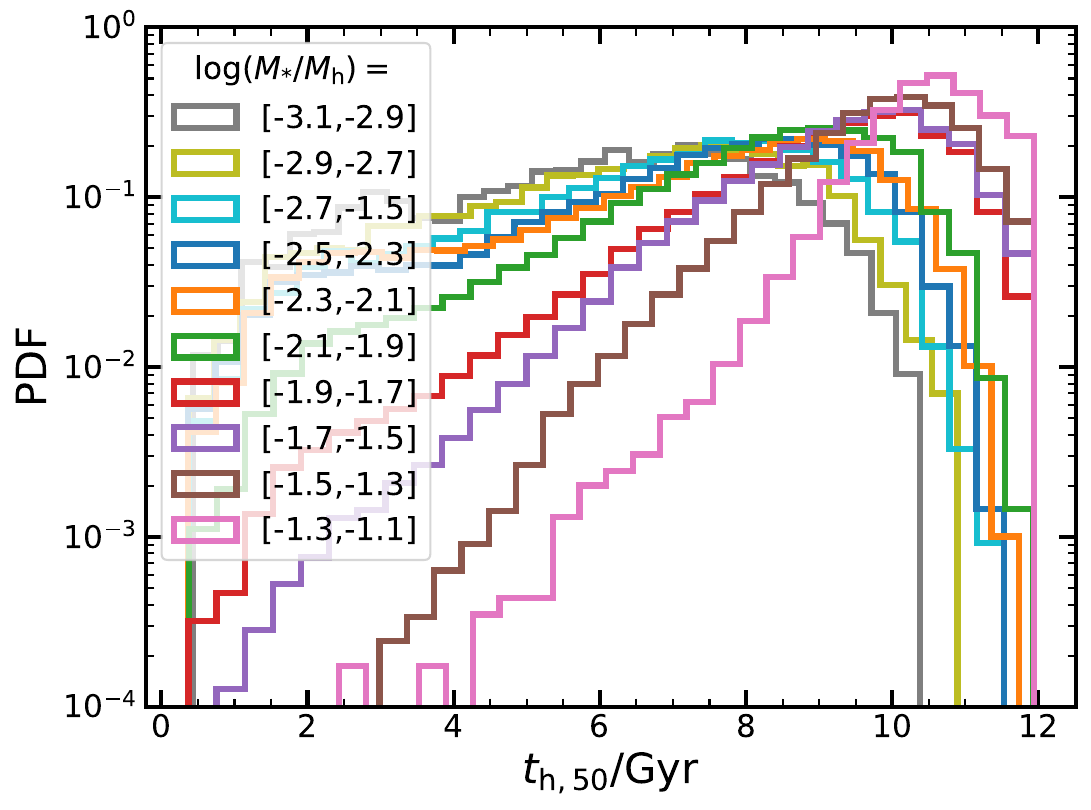}
	\caption{\textbf{Left panel}: Halo assembly time $t_{\mathrm{h,50}}$ as a function of stellar-to-halo mass ratio ($M_*/M_{\mathrm{h}}$) in L-GALAXIES for the blue group (with star-forming central galaxy, SFG) and the red group (with passive central galaxy, PG). Each dot represents a central galaxy/group, plotted for a representative (randomly chosen) 1\% of the sample. Blue and red lines indicate the median values for the blue and red groups, respectively, with shaded regions representing $1\sigma$ uncertainties. \textbf{Right panel}: Normalized $t_{\mathrm{h,50}}$ distribution of central galaxies/groups in each $M_*/M_{\mathrm{h}}$ range, marked by curves in various colors.} \label{t50SHMR}
\end{figure*}

Previous studies have found a positive correlation between the stellar-to-halo mass ratio ($M_*/M_{\mathrm{h}}$) of central galaxies and their halo assembly time ($t_{\mathrm{h,50}}$, defined as the lookback time at which a halo has assembled half of its present-day virial mass), where earlier-formed halos tend to host more massive central galaxies \citep{Wang2011, Lim2016, Tojeiro2017, Bradshaw2020, Correa2020, Wang2023}. This trend is attributed to the fact that earlier-formed halos have more time to accumulate baryons within their central galaxy, making the stellar-to-halo mass ratio a crucial indicator of halo formation chronology. In this work, we illustrate this correlation by plotting $t_{\mathrm{h,50}}$ against $M_*/M_{\mathrm{h}}$ for the blue group (with star-forming central galaxy, SFG, defined as galaxy above Equation (\ref{eqBoundary}) of below Section \ref{DataPreProcessing}) and the red group (with passive central galaxy, PG, defined as galaxy below Equation (\ref{eqBoundary}) of below Section \ref{DataPreProcessing}) in mock data obtained from L-GALAXIES, as shown in the left panel of Figure \ref{t50SHMR}. Notably, there exists a significant systematic difference in the scaling relations between the two groups, particularly at lower $M_*/M_{\mathrm{h}}$ values, up to $\sim$ 4 Gyr. Moreover, both blue and red groups exhibit considerable scatter in $t_{\mathrm{h,50}}$ at a given $M_*/M_{\mathrm{h}}$, especially noticeable in the blue group, potentially introducing inaccuracies and biases in estimating $t_{\mathrm{h,50}}$ solely from $M_*/M_{\mathrm{h}}$. Furthermore, the distribution of $t_{\mathrm{h,50}}$ across various $M_*/M_{\mathrm{h}}$ values is markedly diverse, encompassing a broad range, as further illustrated in the right panel of Figure \ref{t50SHMR}.

Motivated by these limitations and the expectation that other galaxy and group properties may be also correlated with the halo assembly history, we aim to develop a more accurate estimator for $t_{\mathrm{h,50}}$ by incorporating additional galaxy and group properties alongside $M_*/M_{\mathrm{h}}$. To achieve this goal, we initially pre-process the mock data to create unbiased samples with property distributions independent and identical to those observation data. Subsequently, we employ these samples to train our machine-learning models. Finally, we apply these models to observation data to estimate the halo assembly time of each central galaxy. This comprehensive approach enables us to account for various factors to trace halo assembly history, thus enhancing the accuracy of our estimations on halo assembly time.

\subsection{Data pre-processing}\label{DataPreProcessing}

\begin{figure}[htb!] 
	\centering
	\includegraphics[width=0.48\textwidth]{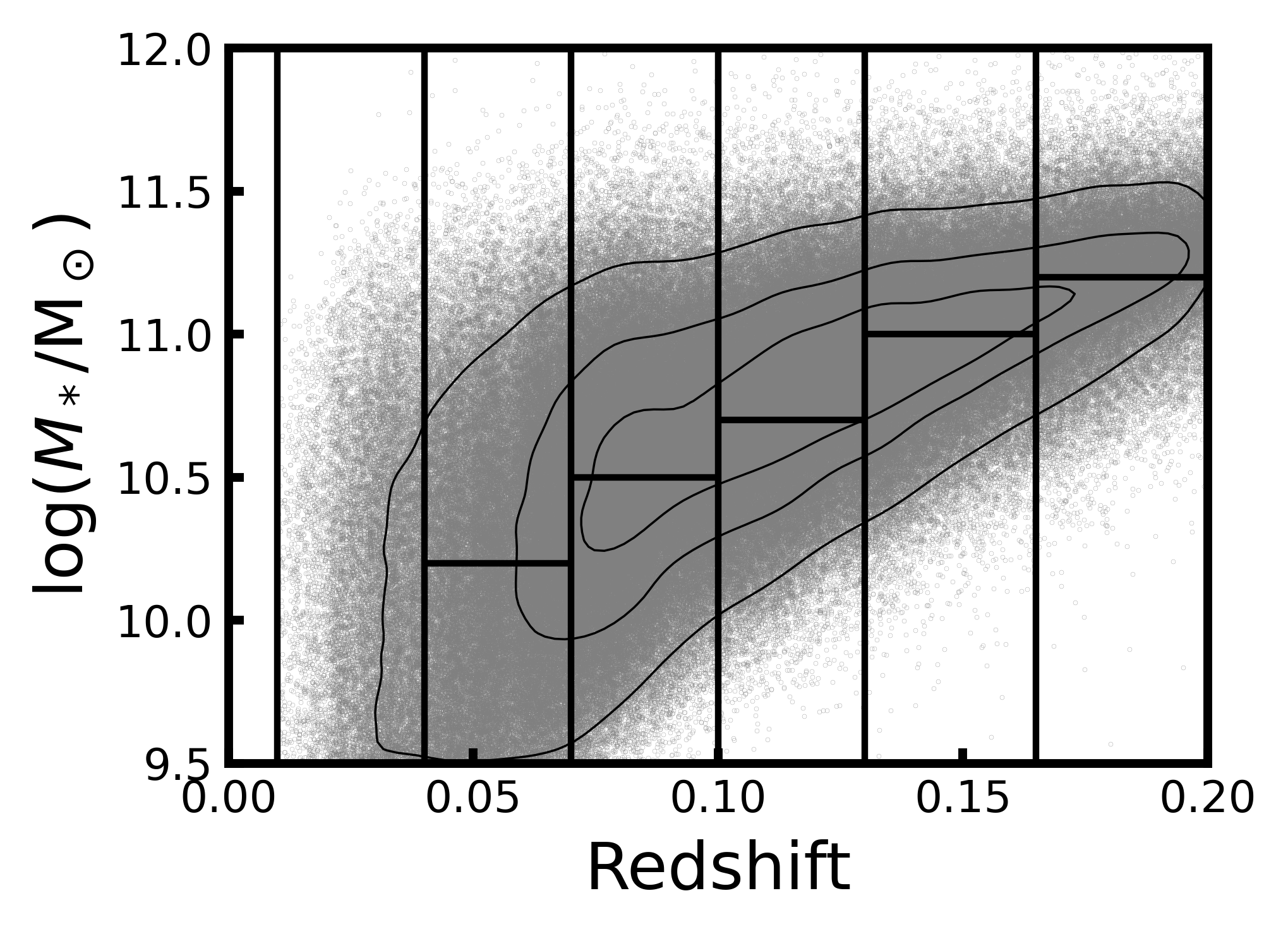}
	\caption{Distribution of central galaxies in stellar mass-group redshift plane for observation data in SDSS. Grey dots stand for galaxies in our observation sample, while black contours show the number density of 90\%, 60\%, and 30\% of the sample. To account for sample selection and incompleteness, we use a similar treatment as in \citet{Zhao2024}, as follows: The vertical lines split central galaxies into six group redshift ranges corresponding to the six snapshots in mocks. For each redshift range, the horizontal line divides the galaxies with $z > 0.04$ into two parts: galaxies above the horizontal line are stellar-mass-complete samples while galaxies below the horizontal line are stellar-mass-incomplete samples. Galaxies with $z < 0.04$ are all stellar-mass-complete samples in this work. The horizontal lines are stellar mass thresholds for each redshift bin, corresponding to $\log(M_*/\mathrm{M_{\odot}})=9.5, 10.2, 10.5, 10.7, 11.0, 11.2$, respectively. \label{MassRedshift}}
\end{figure}

The pre-processing steps for the data are similar to those described in \citet{Zhao2024}, so we simplify and summarize them as follows:

(1) To avoid Malmquist bias and match the snapshot redshift in mocks, we split the whole sample into six subsamples based on different group redshift ranges in observation ($z=0.01\mbox{-}0.04$, $0.04\mbox{-}0.07, 0.07\mbox{-}0.10, 0.10\mbox{-}0.13, 0.13\mbox{-}0.165, 0.165\mbox{-}0.20$, which correspond to the snapshot redshift of $z=0.025612, 0.053316, 0.082661, 0.11378, 0.147548, 0.183387$ in L-GALAXIES). For simplicity, we assign a sample number $i=0,1,2,3,4,5$ to each subsample. To account for sample selection and incompleteness, we use a similar treatment as in \citet{Zhao2024}, as follows: For each redshift bin, we set a stellar mass threshold to split central galaxies into a stellar-mass-complete sample (above stellar mass threshold) and a stellar-mass-incomplete sample (below stellar mass threshold), as shown in Figure \ref{MassRedshift}. The stellar mass thresholds for each redshift bin are $\log(M_*/\mathrm{M_{\odot}})=9.5, 10.2, 10.5, 10.7, 11.0, 11.2$, respectively.

(2) Since star-formation quenching will decouple the co-growth of the dark matter and stellar mass for the central galaxies \citep{Peng2010, Peng2012}, the blue and red centrals are expected to follow different stellar-to-halo mass relations \citep[e.g., illustrated in Figure 2 in][] {Man2019}. This is supported by the weak lensing observations \citep{Mandelbaum2006, Zu2016, Luo2018, Bilicki2021}. Therefore, for each redshift range or snapshot, we divide the groups in the mock data into two samples: blue group (with star-forming central galaxy, SFG) and red group (with passive central galaxy, PG) samples. The SFG/PG are located above/below the boundary for observation data and mock data, respectively:
\begin{equation}
	\log{\mathrm{SFR}}=(0.6+0.02i)\log{M_{*}}-6.9-0.12i,
\end{equation}
\begin{equation}
	\log{\mathrm{SFR}}=(0.9+0.01i)\log{M_{*}}-9.8-0.08i, \label{eqBoundary}
\end{equation}
where $i$ is the sample number, and the physical meanings and units of the quantities are summarized in Table \ref{parameters}.

(3) We assign observational uncertainties (measurement noise) to the mock data to mimic observation data. Following \citet{Zhao2024}, we assign Gaussian noise with mean $\mu_{n}=0$ and standard deviation $\sigma_{n}=\sigma_0=(P84-P16)/2$ to mock data, where $P84$ and $P16$ are the 84th, 16th percentiles of each quantity respectively. This step ensures that the noise level is comparable to the measurement error in observations.

(4) To account for the difference in SFR distribution between the mock and observation data, we adjust the SFR of PG in the mock data. This is done because the SFR of extremely quiescent galaxies is challenging to measure in real observations. Specifically, we assign a new SFR: 
\begin{equation}
	\log{\mathrm{SFR}}=(0.9+0.01i)\log{M_{*}}-10.65-0.08i+\mathcal{N}(0,0.40),
\end{equation}
to the galaxies in mock data below the threshold:
\begin{equation}
	\log{\mathrm{SFR}}=(0.9+0.01i)\log{M_{*}}-10.3-0.08i,
\end{equation}
where $i$ is the sample number and $\mathcal{N}(0,0.40)$ stands for Gaussian distribution with mean $\mu=0$ and standard deviation $\sigma=0.40$. Note that this step solely accounts for the observational effects without introducing artificial effects. Additionally, we have verified that, for PG, SFR minimally impacts halo assembly time estimation (see Figure \ref{FeatureRanking} in Section \ref{Fi}).

(5) To reduce the bias between the mock and observation data, we normalize the input quantities before training by using the following equation:
\begin{equation}
	X=\frac{X-\mu_{X}}{\sigma_{X}},
\end{equation}
where $X$ represents the SFR, age, color indices, and $M_*/M_{\mathrm{h}}$. For each blue/red group subsample at a given redshift, we compute the mean $\mu_{X}$ and standard deviation $\sigma_{X}$ of $X$ and then normalize $X$ accordingly. We note that for mass properties such as $M_{*}$, $M_{\mathrm{h}}$, and $M_{\mathrm{tot}}$, we do not perform this step, because for the stellar-mass-complete samples, as these properties in both the mocks and observations follow similar distribution function (i.e., halo mass function, stellar mass function, see \citet{Zhao2024} for more details). For the stellar-mass-incomplete samples, we resample the mock data to mimic the $M_{\mathrm{h}}$ and $M_{*}$ distributions in observation samples below the $M_{*}$ threshold (as defined in Figure \ref{MassRedshift}) within each redshift range.

\subsection{Machine learning method}

We employ the {\ttfamily XGBoost} algorithm \citep{Chen2016} to construct regressors for estimating the $t_{\mathrm{h,50}}$ of the blue/red groups in each redshift range. This ensemble learning algorithm is based on gradient-boosted trees and is known for its simplicity and effectiveness.

Specifically, we divide the blue/red groups in a given redshift range of the mock data into train, test, and validation samples in an 8:1:1 ratio. We then train and test the regression models for the stellar-mass-complete samples using several galaxy properties as input quantities, as summarized in Table \ref{parameters}. For the stellar-mass-incomplete galaxy samples, we do not use their total stellar mass due to the biased observation selection. To test whether each property contributes to the halo assembly history, we also add uniform random values as input quantiles to the models. The output is $t_{\mathrm{h,50}}$ of each group. To implement {\ttfamily XGBoost} into our mock data, we utilize the {\ttfamily scikit-learn} of $Python$ package with the {\ttfamily n\_estimators} hyper-parameter set to 100 (which corresponds to the number of trees in the forest). We have also tuned other hyper-parameters such as {\ttfamily max\_features} and {\ttfamily max\_depth} for tests, but the results are relatively insensitive to these changes.

\section{Results}\label{result}

\subsection{Model regression}

\begin{figure*}[htb!] 
	\centering
	\includegraphics[width=0.48\textwidth]{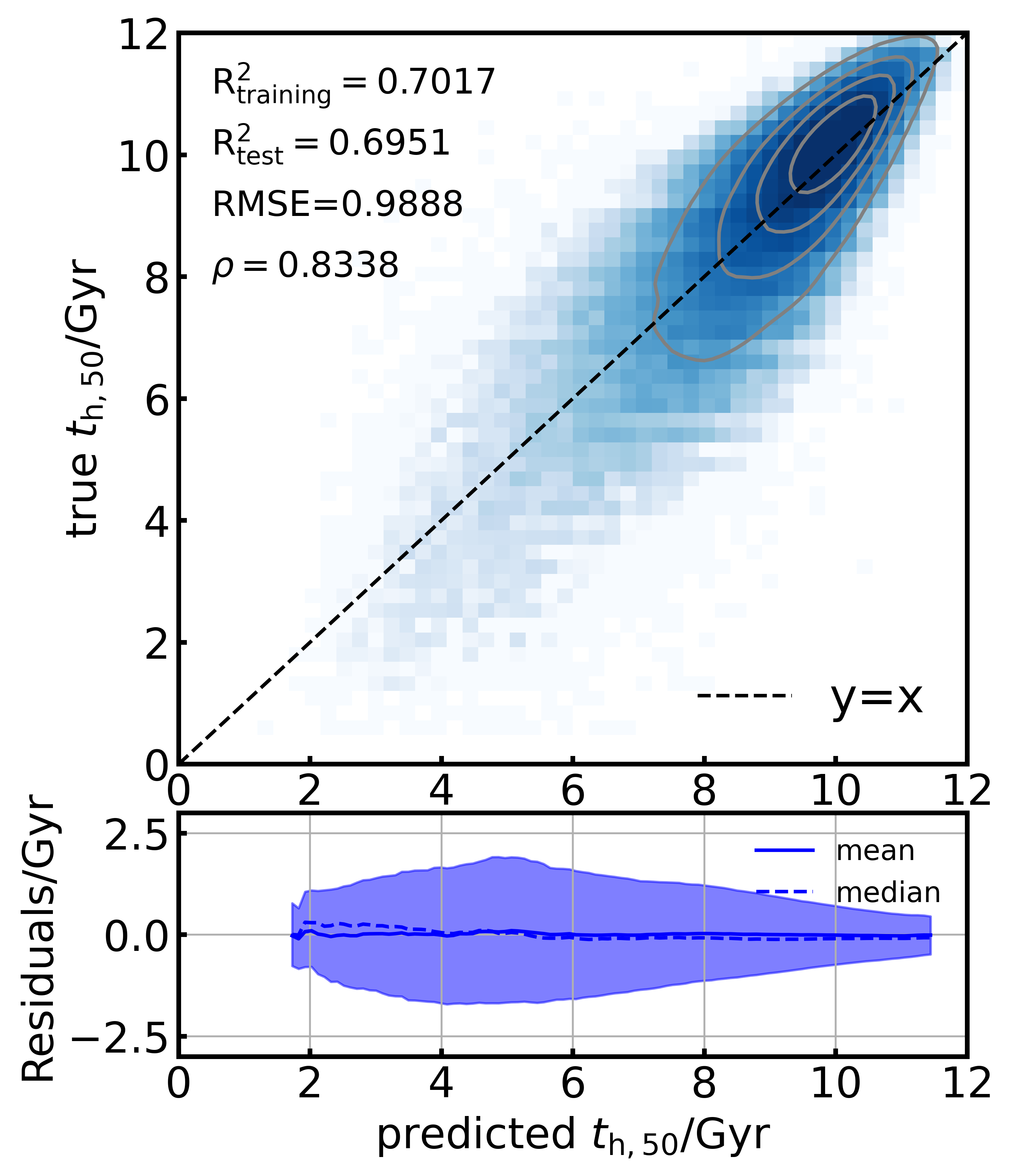}
	\includegraphics[width=0.48\textwidth]{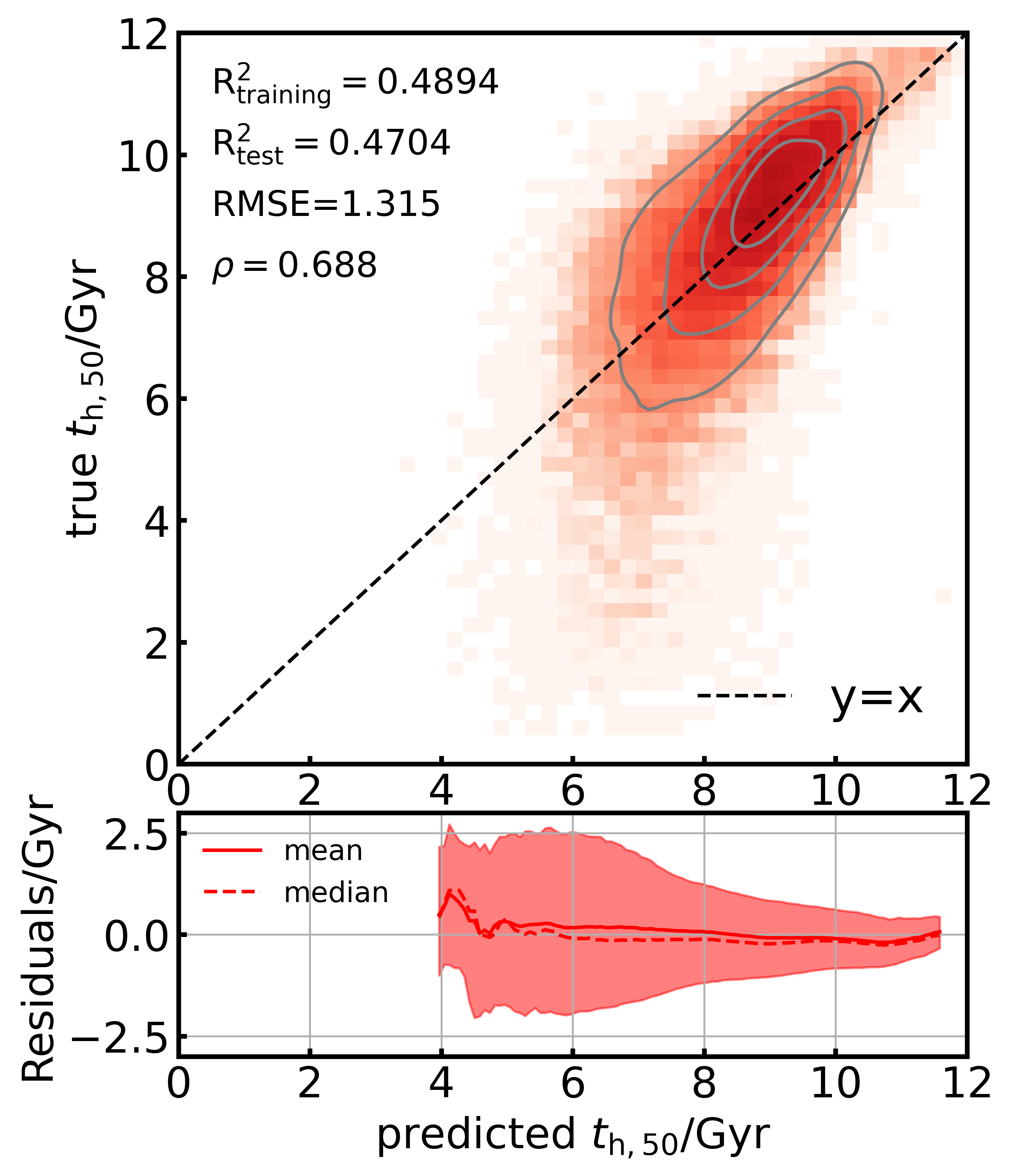}
	\caption{\textbf{Top panels}: Two-dimensional representations of the true values (y-axis) versus predicted (x-axis) values of $t_{\mathrm{h,50}}$ for blue group (left) and red group (right) mock samples using machine-learning regressor on the test sample of the stellar-mass-complete sample models within 0.01<$z$<0.04. The color coding is the number density of the galaxies in logarithmic scales, and the grey contours cover 20\%, 40\%, 60\%, and 80\% of the test sample. The dashed line indicates the 1:1 relation. The assessment parameters are marked at the top left of the top panels: training score ($R^2_{\mathrm{training}}$), test score ($R^2_{\mathrm{test}}$), root-of-mean-square-error (RMSE) and Pearson correlation coefficient $\rho$. Models perform better if they show lower RMSE and higher $\rho$. Note that for red group samples, those with $t_{\mathrm{h,50}}$ values below 5 Gyr only account for less than 5\% of the total mock sample, and this holds across all snapshots. \textbf{Bottom panels}: The mean residuals (solid lines) and the median residuals (dashed lines) as a function of predicted $t_{\mathrm{h,50}}$ values for the blue group (left) and the red group (right) mock samples, with shaded regions of 16th to 84th quantiles of variation. Note that for a given predicted $t_{\mathrm{h,50}}$, the true $t_{\mathrm{h,50}}$ distribution can be asymmetry. Specifically, at the high predicted $t_{\mathrm{h,50}}$ range, although the contour lines exhibit some deviation from the one-to-one relation (top panels), the residuals of medians and means reveal no significant biases (bottom panels).}
	\label{Regression}
\end{figure*}

Using mock data, we present the performance of regression models on test samples for the stellar-mass-complete samples within 0.01<$z$<0.04 in Figure \ref{Regression}. The assessment parameters, displayed in the top-left corner of each top panel, validate the effectiveness of our regression models on the test sample. Specifically, the Pearson correlation coefficient is $\sim$0.8338(0.688), and the root-of-mean-square-error (RMSE) is $\sim$0.9888(1.315) Gyr for blue(red) group samples. Considering all the samples including blue and red groups, the average RMSE is $\sim$1.09 Gyr. We employ RMSE as the uncertainty measure for $t_{\mathrm{h,50}}$ estimation. The close alignment between the training and test scores suggests that our models effectively avoid overfitting. The mean and median residuals of $t_{\mathrm{h,50}}$ value as a function of the predicted $t_{\mathrm{h,50}}$ value on the test sample are depicted in the bottom panels. Note that for a given predicted $t_{\mathrm{h,50}}$, the true $t_{\mathrm{h,50}}$ distribution can be asymmetry. Specifically, at the high predicted $t_{\mathrm{h,50}}$ range, although the contour lines exhibit some deviation from the one-to-one relation (top panels of Figure \ref{Regression}), the residuals of medians and means reveal no significant biases (bottom panels of Figure \ref{Regression}). Thus, the models successfully recover the $t_{\mathrm{h,50}}$, with the average uncertainty of $\sim$ 0.99 Gyr for the blue group and $\sim$ 1.32 Gyr for the red group.

However, the models underperform for red groups that assembled relatively later (with lower $t_{\mathrm{h,50}}$ values), as depicted in the right panels of Figure \ref{Regression}. A possible physical explanation is that a significant fraction of these low-$t_{\mathrm{h,50}}$ samples may have recently undergone a halo merger, while the galaxies themselves have not merged yet. Consequently, the $t_{\mathrm{h,50}}$ calculated based on the present-day halo mass is lower, but the galaxies trace back to the properties before the merger, causing the predicted $t_{\mathrm{h,50}}$ to appear higher than its intrinsic value. It is worth noting that for red group samples, those with $t_{\mathrm{h,50}}$ values below 5 Gyr only account for less than 5\% of the total mock sample, and this holds true across all snapshots. Attempts to include non-observable or hard-to-observe properties have scarcely improved model performance for this small population. We intend to further investigate this phenomenon in future studies.

\begin{figure*}[htb] 
	\centering
	\includegraphics[width=0.48\textwidth]{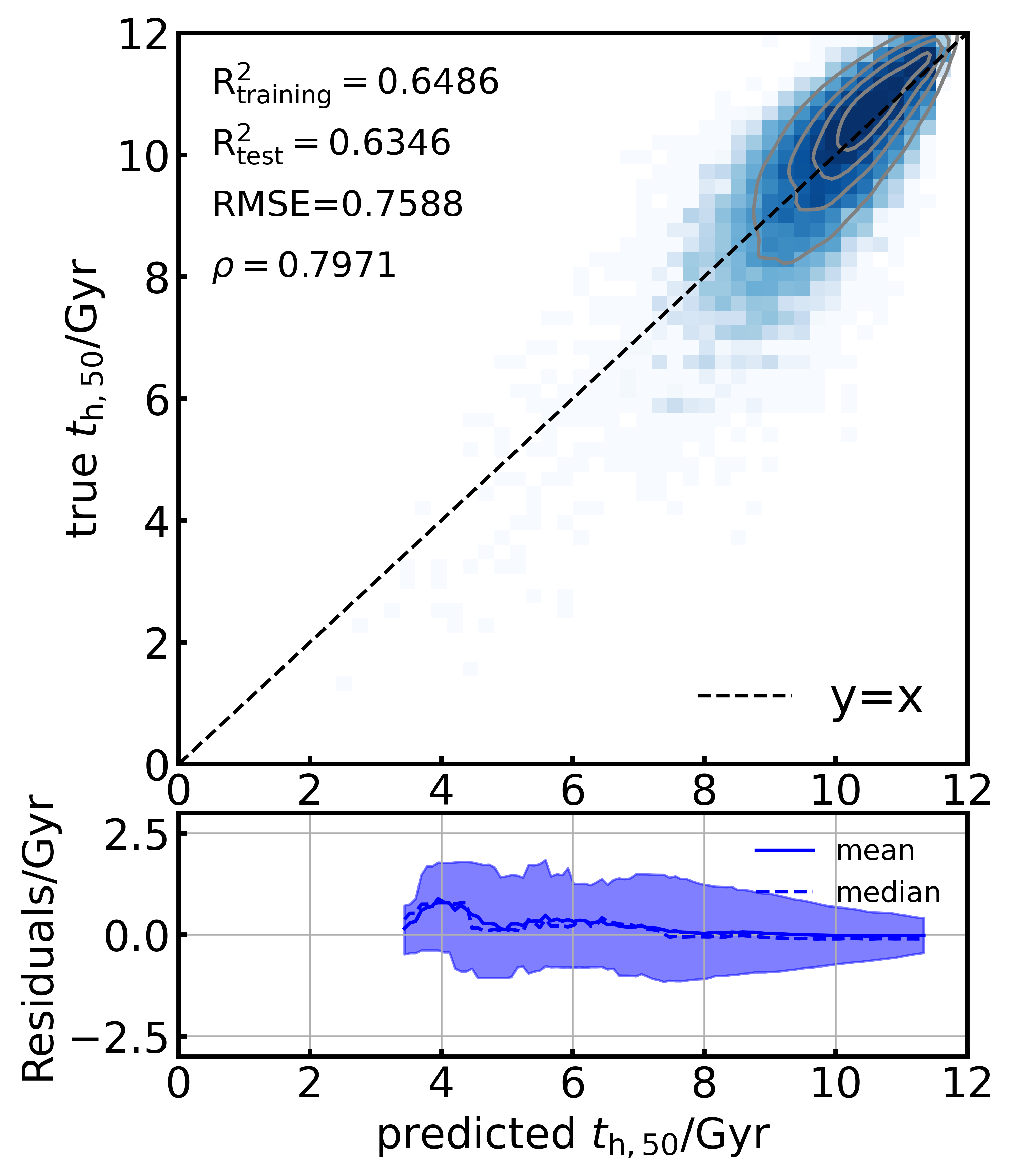}
	\includegraphics[width=0.48\textwidth]{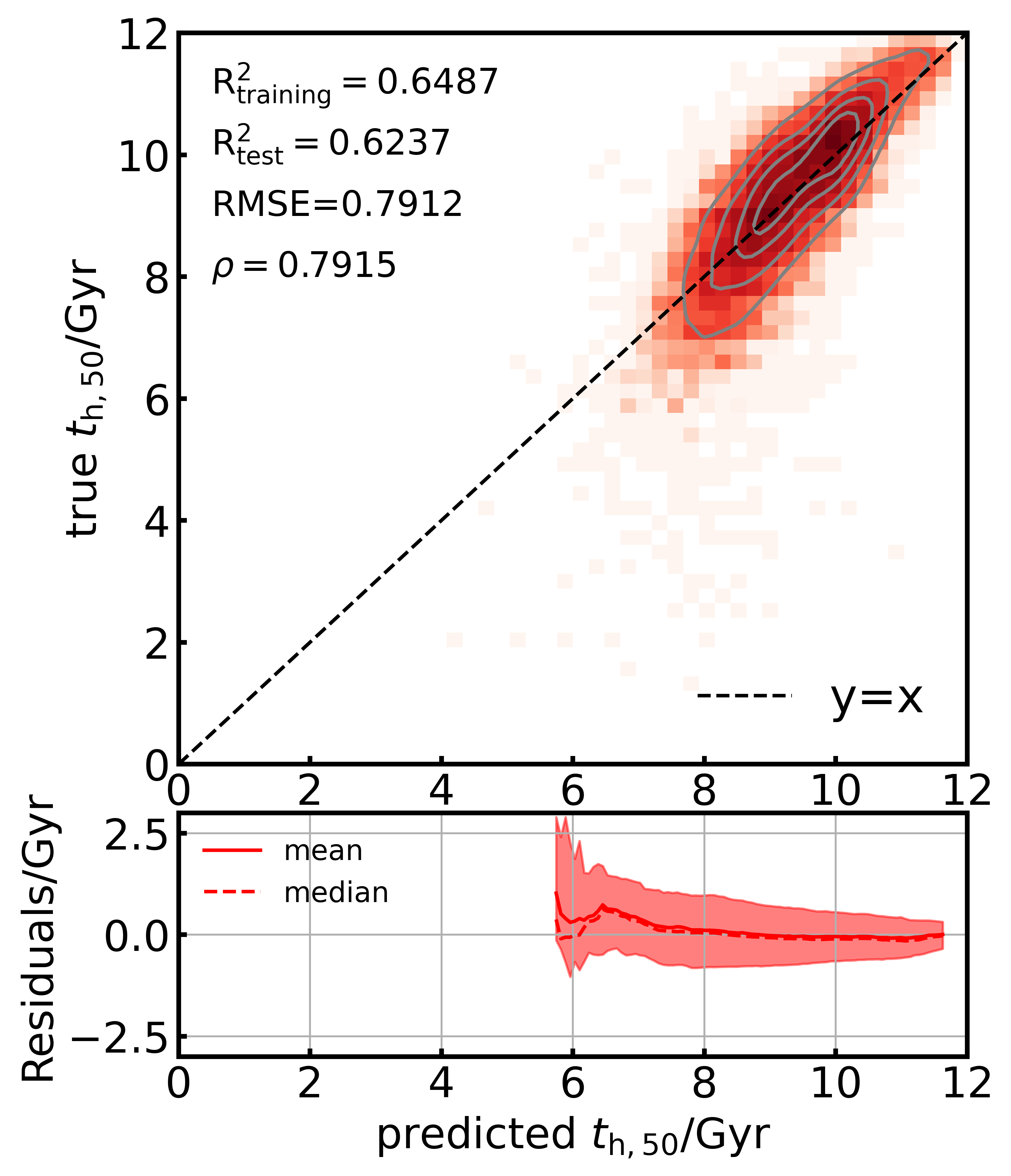}
	\caption{Similar to Figure \ref{Regression}, but for the test sample of the stellar-mass-incomplete sample models within 0.04<$z$<0.07.}
	\label{Regression004007inc}
\end{figure*}

\begin{figure*}[htb]
	\centering
	\includegraphics[width=0.48\textwidth]{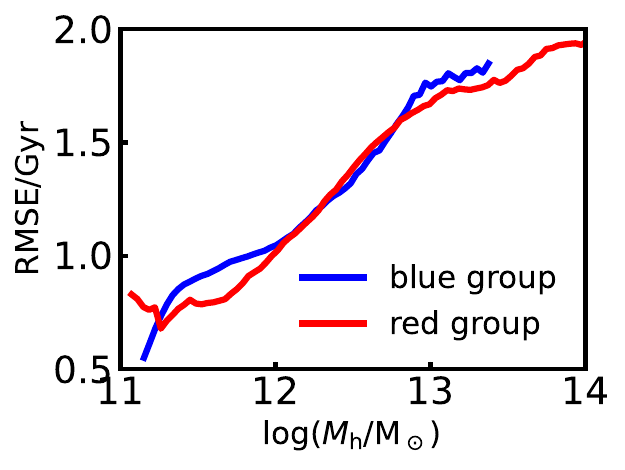}
	\includegraphics[width=0.48\textwidth]{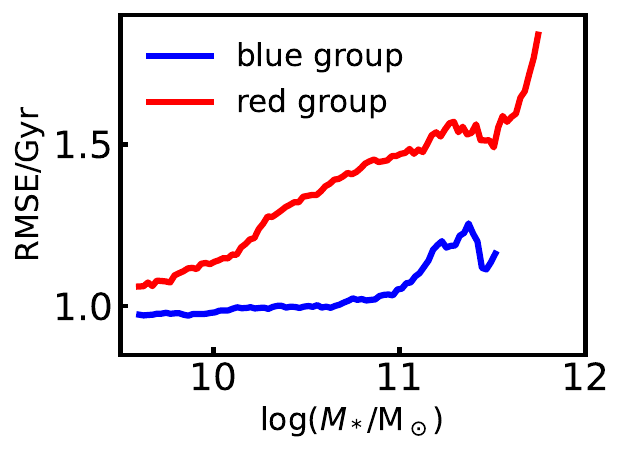}
	\caption{Root-of-mean-square-error (RMSE) as a function of halo mass (left panel) and stellar mass of the central galaxy (right panel) for the test sample of the stellar-mass-complete sample models within 0.01<$z$<0.04. The blue and red curves represent the median RMSE for the blue and red groups, respectively.}
	\label{M-RMSE}
\end{figure*}

For the stellar-mass-incomplete samples, we present the performance of regression models on the test samples within 0.04<$z$<0.07 in Figure \ref{Regression004007inc}. Compared to the models for the stellar-mass-complete samples within the same redshift range (refer to Figure \ref{Regression004007com} in the Appendix), the distributions for the stellar-mass-incomplete samples align more closely with the 1:1 relation in the top panels, as indicated by lower values of RMSE ($\sim$0.7588 for blue group and $\sim$0.7912 for red group). This trend is likely due to the differing halo mass and stellar mass distributions in the mock data for stellar-mass-complete and -incomplete samples, as well as mass-dependent RMSE in $t_{\mathrm{h,50}}$ estimation. As depicted in the left panel of Figure \ref{M-RMSE}, the RMSE exhibits a strong positive relation with halo mass. This trend suggests that estimating $t_{\mathrm{h,50}}$ for more massive halos is generally more challenging, primarily due to their complex assembly histories (e.g., involving more mergers). On average, more massive halos tend to assemble later with more complex assembly histories, resulting in larger RMSE values. At a given halo mass, the RMSE values do not significantly vary between the blue and red groups because halo mass assembly occurs regardless of the star formation status of its central galaxy. As shown in the right panel of Figure \ref{M-RMSE}, the RMSE generally increases with stellar mass of the central galaxy. At a given stellar mass of the central galaxy, the RMSE for the red group exceeds that of the blue group because the halo mass of the red group is typically larger than that of the blue group \citep{Mandelbaum2006, Zu2016, Luo2018, Bilicki2021}. Therefore, it implies a more complex assembly history of the red group at a given stellar mass. Furthermore, it is important to note that most central galaxies in the blue groups have $\log(M_{*}/\mathrm{M_{\odot}})<11.0$ and $\log(M_{\mathrm{h}}/\mathrm{M_{\odot}})<12.5$, resulting in an average RMSE for the blue group of less than $\sim$ 1 Gyr. Additionally, the RMSE for the blue group shows a relatively steeper increase at around $\log(M_{*}/\mathrm{M_{\odot}}) \sim 11.0$, aligning with the ``knee'' in the stellar-to-halo mass relation \citep[e.g.,][]{Guo2010, Moster2013}. Beyond this stellar mass, a given $M_{*}$ corresponds to a wide range of $M_{\mathrm{h}}$, leading to a broad RMSE range as depicted in the left panel of Figure \ref{M-RMSE}. These trends in Figure \ref{M-RMSE} hold for both stellar-mass-complete samples and stellar-mass-incomplete samples across all snapshots in mock data.

\subsection{Feature importance}\label{Fi}

\begin{figure*}[htb!]
	\centering
	\includegraphics[width=0.48\textwidth,height=0.2\linewidth]{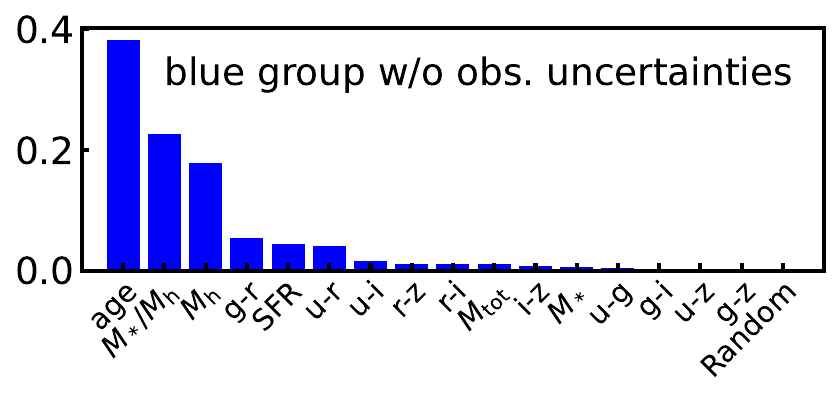}
	\includegraphics[width=0.48\textwidth,height=0.2\linewidth]{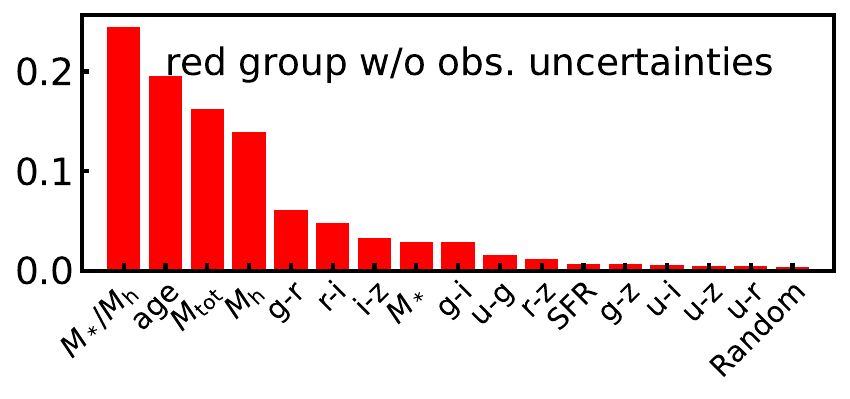}
	\includegraphics[width=0.48\textwidth,height=0.2\linewidth]{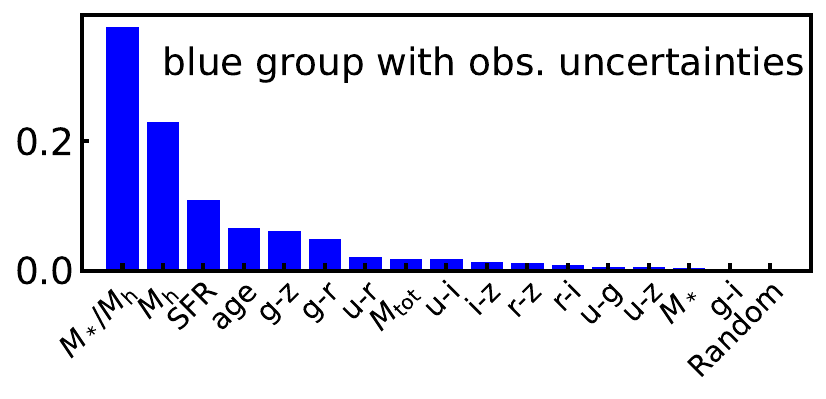}
	\includegraphics[width=0.48\textwidth,height=0.2\linewidth]{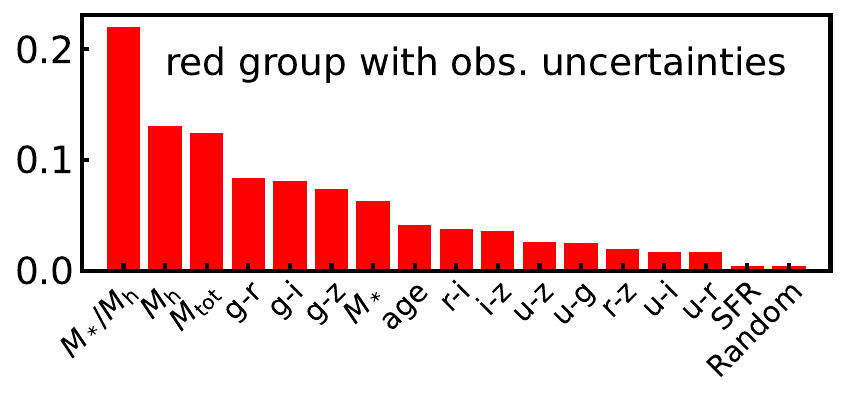}
	\caption{The relative importance of all input quantities (labeled on the x-axis) employed for the training of the machine-learning regressors in mock data of the blue group (left panels) and the red group (right panels) samples of the stellar-mass-complete sample models within 0.01<$z$<0.04, without observational uncertainties (top panels) or with observational uncertainties (bottom panels). \label{FeatureRanking}}
\end{figure*}

Given that the properties of central galaxies can be used to infer $t_{\mathrm{h,50}}$ with reasonable accuracy, it is possible to determine which specific properties contain the information that maximizes the predictive capability of the models. We expect that at a given halo mass, central galaxies in early-assembled halos on average have a longer time to form stars and accumulate much more stellar mass, linking $t_{\mathrm{h,50}}$ closely with stellar age, stellar-to-halo mass ratio, and halo mass.

In the bottom panels of Figure \ref{FeatureRanking}, we illustrate the relative feature importance of all input quantities employed for training the machine-learning regressors for the blue group (left panel) and the red group (right panel) mock samples on the test sample of the stellar-mass-complete sample models with observational uncertainties within 0.01<$z$<0.04. Relative feature importance is a widely used approach to quantifying and interpreting the contribution of different input features to the predicted variable. In \texttt{scikit-learn}, relative importance is ranked based on the Gini importance \citep{Pedregosa2011}. As depicted, all 16 quantities we adopted show greater importance than random numbers, suggesting correlations with $t_{\mathrm{h,50}}$.

In the top panels of Figure \ref{FeatureRanking}, we also display the relative feature importance of trained noise-free models, to disentangle the specific role of each quantity in estimating $t_{\mathrm{h,50}}$. This is crucial since varying levels of observational uncertainties (measurement noise) can obscure the information encoded in galaxy properties and affect their contributions to the models. These rankings confirm our expectations: In the noise-free models, roughly, the stellar age and $M_*/M_{\mathrm{h}}$ are the dominant properties, followed by halo mass and total mass, while most color indices are less significant. These results provide a clear picture of the model prediction that the correlations between $t_{\mathrm{h,50}}$ and stellar age, and between $t_{\mathrm{h,50}}$ and $M_*/M_{\mathrm{h}}$, are stronger, with a secondary dependence on halo mass. These correlations can be attributed to the fact that at a given halo mass at $z=0$, different $t_{\mathrm{h,50}}$ partly determines the occurrence time of the main star formation activities in central galaxies. However, when observational uncertainties are assigned to the mock data (bottom panels in Figure \ref{FeatureRanking}), the relative importance of stellar age is significantly reduced due to its measurement uncertainties. Furthermore, some colors, such as $g-r$, usually rank higher than others due to their heightened sensitivity to star formation rates and stellar age. For instance, previous studies utilized $g-r$ color to categorize galaxies into distinct sequences or morphologies based on bimodal distributions in color-magnitude or color-mass relations \citep[e.g.,][]{Strateva2001, Bell2003, Baldry2004, Baldry2006, Blanton2009}.

It is important to recognize that lower-ranked properties in feature importance analysis do not necessarily indicate a negligible correlation with $t_{\mathrm{h,50}}$. Instead, these relationships might be overshadowed or encapsulated within other higher-ranked properties, such as those at the forefront. i.e., they contribute minimally to additional information in model regression. It is crucial to note that the correlation may not appear weak when considering it separately with $t_{\mathrm{h,50}}$.

\subsection{Apply models to observation data}

With careful calibrations of individual observable quantities, we estimate the $t_{\mathrm{h,50}}$ of observation data based on trained regression models. In this work, we focus on central galaxies in the SDSS DR7 sample within 0.01<$z$<0.20 and $\log (M_{*}/\mathrm{M_\odot})>9.5$, matched with the \citet{Yang2007} group catalog and the \citet{Zhao2024} halo mass catalog. For central galaxies in our sample with valid measurements of observable quantities, we can approximately infer their $t_{\mathrm{h,50}}$ of their host halo. For the stellar-mass-complete/-incomplete blue/red group samples at various redshift ranges, we apply the regression models trained by corresponding mock samples to observation data.

Note that some central galaxies lack age measurements in the observation data. To avoid potential systematic biases stemming from this absence of data, we also develop alternative models that exclude age as an input to estimate $t_\mathrm{h,50}$ for these groups. These models are also trained by calibrated mock data similar to fiducial models. The new models demonstrate similar performance (with RMSE of $\sim$ 1.01 Gyr for blue groups and $\sim$ 1.32 Gyr for red groups within 0.01<$z$<0.04) in the mocks to our fiducial models including age as an input.

\subsection{Distributions of halo assembly time}

\begin{figure*}[htb!]
	\centering
	\includegraphics[width=1\textwidth]{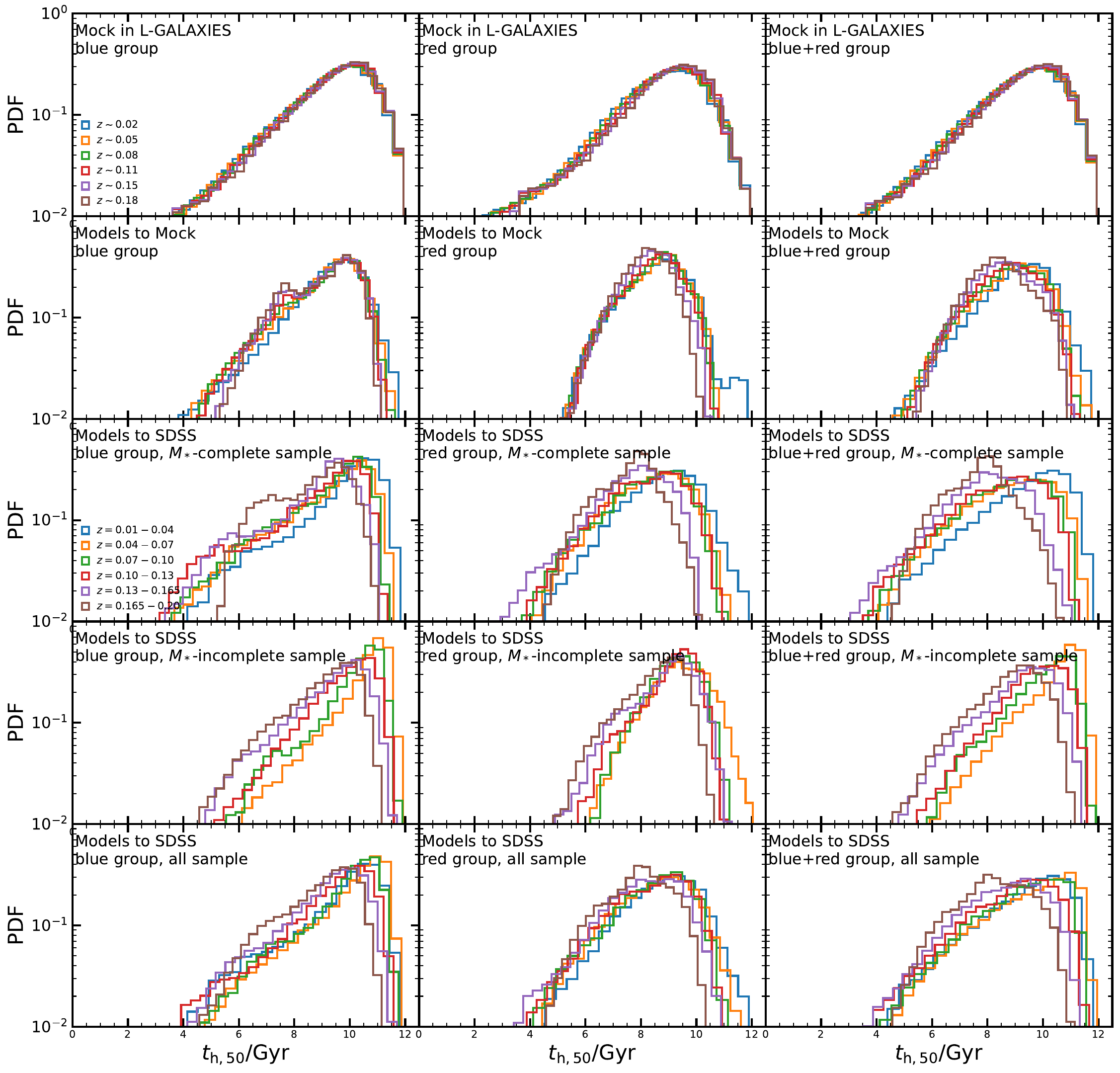}
	\caption{Distributions of $t_{\mathrm{h,50}}$ for various snapshot/redshift ranges for mock data in L-GALAXIES (\textbf{top row}), apply trained models to mock data (\textbf{second row}), apply trained models to stellar-mass-complete samples, as defined in Figure \ref{MassRedshift} (\textbf{third row}), apply trained models to stellar-mass-incomplete samples (\textbf{fourth row}), and to all samples of this work in SDSS (\textbf{bottom row}). The left, middle, and right columns are for the blue group, red group, and blue+red group, respectively. Each curve of different colors corresponds different snapshot or redshift range indicated in the legends. Note that for the red group, samples whose $t_{\mathrm{h,50}}$ is not well reproduced (i.e. $t_{\mathrm{h,50}}<5$ Gyr), only account for less than 5\% of the total mock sample, and this holds across all snapshots. In the top row, the distributions of $t_{\mathrm{h,50}}$ in the mock data using a stellar-mass-complete sample with $\log (M_*/\mathrm{M_\odot})>9.5$ at each snapshot. The mock data used for training (second row) adhere to the same selection criteria as those for the SDSS (as depicted in Figure \ref{MassRedshift}). In the third, fourth, and fifth rows, the distributions of $t_{\mathrm{h,50}}$ in SDSS tend to shift slightly to the lower values with increasing redshift, primarily because SDSS groups at lower redshifts on average contain a significantly higher number of low-mass galaxies, corresponding to low-mass halos. These halos generally assembled earlier with higher $t_{\mathrm{h,50}}$ values than those at higher redshifts.}
	\label{t50eachz}
\end{figure*}

We present the $t_{\mathrm{h,50}}$ distributions for various snapshot/redshift ranges for mock data in L-GALAXIES (top row), apply trained models to mock data (second row), apply trained models to stellar-mass-complete samples, as defined in Figure \ref{MassRedshift} (third row), apply trained models to stellar-mass-incomplete samples (fourth row), and to all samples of this work in SDSS (bottom row) in Figure \ref{t50eachz}, to test the model validity and the reliability of our methods. Notably, the $t_{\mathrm{h,50}}$ distributions at different redshift bins exhibit no discernible cosmic evolution in the mock data using stellar-mass-complete samples with $\log (M_*/\mathrm{M_\odot})>9.5$. This is expected due to the relatively narrow redshift range explored here, $0<z<0.2$. However, upon applying the trained models to SDSS samples, the $t_{\mathrm{h,50}}$ distributions exhibit some redshift dependence and variations at both low and high tails of $t_{\mathrm{h,50}}$.

For SDSS groups, the distributions of $t_{\mathrm{h,50}}$ are slightly different at different redshifts, as shown in each panel in the third, fourth, and fifth rows of Figure \ref{t50eachz}. Compared to higher redshift, SDSS groups at lower redshifts on average contain a significantly higher number of low-mass galaxies (no matter for stellar-mass-complete or stellar-mass-incomplete samples, see Figure \ref{MassRedshift}). The host halos of these low-mass galaxies, statistically corresponding to low-mass halos, generally assembled earlier with higher $t_{\mathrm{h,50}}$ values (according to hierarchical halo assembly). Therefore, there is a general trend where the distributions of $t_{\mathrm{h,50}}$ tend to shift slightly to the lower values with increasing redshift for each panel in the third, fourth, and fifth rows of Figure \ref{t50eachz}. Additionally, the mock data used for training, which vary across different redshifts, have been assigned different noise levels to mimic measurement uncertainties. This approach also leads to systematic differences in the models and the derived distribution of $t_{\mathrm{h,50}}$ for SDSS groups at various redshifts.

For SDSS groups, at each redshift, the distributions of $t_{\mathrm{h,50}}$ for stellar-mass-complete samples (third row of Figure \ref{t50eachz}) show more significant low-$t_{\mathrm{h,50}}$ tails than those for stellar-mass-incomplete samples (fourth row of Figure \ref{t50eachz}). This is again, primarily due to the higher average stellar mass for the stellar-mass-complete samples (as shown in Figure \ref{MassRedshift}), which statistically corresponds to higher-mass halos that generally assembled later with lower $t_{\mathrm{h,50}}$ values (according to hierarchical halo assembly).

The estimation of $t_{\mathrm{h,50}}$ for groups in the ``tails" is relatively challenging: Groups with lower $t_{\mathrm{h,50}}$ values (i.e., in the low $t_{\mathrm{h,50}}$ tails) often represent unsettled systems, such as those undergoing recent mergers, making it particularly difficult to infer halo properties from galaxy properties. Conversely, groups with higher $t_{\mathrm{h,50}}$ values (i.e., in the high $t_{\mathrm{h,50}}$ tails) typically consist of halos and galaxies that assembled very early, which also complicates inferring $t_{\mathrm{h,50}}$ solely based on galaxy properties at $z\sim0$. However, it is worth noting that such samples in the ``tails" constitute only a small fraction of the entire dataset.

For the stellar-mass-complete blue groups, the $t_{\mathrm{h,50}}$ distribution broadly reproduces the shapes in mocks. While for red groups, the caveats of our model prediction are revealed: when applied the trained models to mock data, the shape of $t_{\mathrm{h,50}}$ distributions cannot be totally well reproduced, especially lacking low $t_{\mathrm{h,50}}$ ones (samples of $t_{\mathrm{h,50}}<5$ Gyr account for less than 5\% of the total mock sample). This discrepancy can be attributed to the biased regression in models for red groups, as shown in the right panels of Figure \ref{Regression}, indicating that the galaxy properties in red groups assembled relatively recently fail to accurately trace back to the halo properties. For the stellar-mass-incomplete samples, the $t_{\mathrm{h,50}}$ distributions exhibit sharp-tip shapes, possibly due to the incompleteness of quantities used. For both blue and red groups, the $t_{\mathrm{h,50}}$ distributions at higher redshifts show progressively larger deviations from mocks, likely due to the increasingly observational uncertainties in group findings and property estimations. Moreover, there may be a lot of non-negligible information on $t_{\mathrm{h,50}}$ hidden in galaxies or dark matter halo properties that we do not adopt. Overall, the derived SDSS $t_{\mathrm{h,50}}$ distributions are in reasonable agreement with mocks, in particular for blue groups and groups at lower redshift, which hence strongly supports our methods.

\section{Discussions}\label{discussion}

\subsection{The selection of properties}\label{WhyProperties}
 
In this work, we aim to identify present-day galaxy and group properties that can be traced back to halo assembly time and obtain the most accurate estimation. In the model training phase, we meticulously select galaxy and group properties that are relatively reliable in the mocks, and are tightly correlated to assembly history of both galaxy and halo. The selection criteria of galaxy and group properties have been clarified in Section \ref{Ggp}. It is important to acknowledge that while our model offers estimations of $t_{\mathrm{h,50}}$, it does not achieve perfect regression. This limitation suggests the possible existence of other properties that are linked to the halo assembly history yet remain unobservable with current tools. Additionally, random processes may also influence the determination of $t_{\mathrm{h,50}}$.

Previous studies have identified several galaxy properties that are associated with halo assembly history and can thus act as indicators of formation time, such as magnitude gap \citep[e.g.,][]{Dariush2010, Farahi2020}, halo concentration \citep[e.g.,][]{Deason2013}, and environmental factors \citep[e.g.,][]{Harker2006, Wechsler2006, Tojeiro2017}, etc. Initially, we have attempted to incorporate a broad array of input quantities into the mock data, spanning central galaxy properties (such as cold gas mass, hot gas mass, gas-phase metallicity, stellar metallicity, gas spin, stellar spin, black hole mass, bulge size, and disk size), galaxy group properties (such as group richness, magnitude gaps, and stellar mass gaps), and environmental indicators (such as number density and distance to the 5th distant central galaxy). However, we find that augmenting these quantities did not significantly enhance the accuracy of our $t_{\mathrm{h,50}}$ estimations. Moreover, some properties pose challenges in observation, exhibit small matched samples with the SDSS catalog, or are plagued by significant measurement uncertainties. Although colors play a minor role in model regression, they demonstrate relatively high reliability in L-GALAXIES and align well with observations after meticulous calibration. Consequently, we strike a balance between the availability and significance of galaxy properties, ultimately opting to include only 16 quantities in this work.

It may be argued that the structure, morphology, and gas properties are also pertinent to galaxy and halo assembly history and should thus be incorporated as model inputs. However, semi-analytic models struggle to properly account for these aspects (even hydrodynamical simulations encounter similar challenges). Furthermore, different hydrodynamical simulations and semi-analytic models yield disparate outcomes regarding structure, morphology, and gas properties due to variations in baryon recipes, such as feedback mechanisms. Moreover, random mergers or interactions further obscure intrinsic correlations between halo and galaxy structure, morphology, and gas properties. Therefore, we decide to not incorporate these variables in our model training. Another consideration is that we plan to explore the correlation between halo properties and galaxy properties (including structure, morphology, and gas properties) using SDSS data in future studies. To minimize circularity (though it remains inevitable), we prioritize leveraging the most reliable quantities reproduced by the mock data compared to observations.

\subsection{Why L-GALAXIES?}

In this work, we utilize mock data in L-GALAXIES semi-analytic model as the training dataset. One may argue that hydrodynamical simulations may seem to generate more accurate mocks in $t_{\mathrm{h,50}}$ estimations compared to semi-analytic models. In general, the hydrodynamical simulations are more physical than semi-analytic models. However, the semi-analytic models facilitate easier calibration and tuning of parameters to align with observations, making them particularly effective for statistical analyses. As mentioned in Section \ref{LG}, L-GALAXIES has been meticulously calibrated and tuned to reproduce fundamental observed properties such as the stellar mass function, star formation main sequence, stellar-to-halo mass relations, and quenched fraction across various redshifts, which impose robust constraints on star formation and halo assembly histories. For instance, L-GALAXIES successfully aligns with weak lensing measurements for stellar-to-halo mass relations in red/blue groups \citep{Zhao2024}, in contrast to IllustrisTNG \citep{Nelson2018}, which has shown limitations in this respect \citep{Zhang2022}, although SIMBA \citep{Dave2019} demonstrates strong performance in these areas \citep{Cui2021}.

Therefore, for hydrodynamical simulations and semi-analytic models employing diverse recipes, if they can accurately reproduce observed stellar mass functions and quenched fractions, it is reasonable to expect they will also reflect similar star formation histories and halo assembly histories. On the other hand, for those properties with large observational uncertainties, such as structure, morphology, size and gas properties, different hydrodynamical simulations and semi-analytic models may yield divergent results. Consequently, we have opted not to incorporate these properties as training inputs, as discussed in Section \ref{WhyProperties}.

The primary scientific goal of this paper is to highlight the feasibility of utilizing observable properties to infer and constrain halo assembly history, rather than assessing the correctness of physical models underlying these simulations. Therefore, our preferred approach involves using a semi-analytic model or hydrodynamical simulation that best reproduces observed star formation and halo assembly histories, despite potential inaccuracies in the underlying physics. We plan to test and validate these results across different hydrodynamical simulations and other semi-analytic models, such as IllustrisTNG, EAGLE \citep{Schaye2015}, and SIMBA, which are involved in more related baryonic processes, if large volumes are not required (to minimize cosmic variance).

\subsection{Model limitations}

It is important to clarify that our models are not designed to provide perfectly accurate measurements of the $t_{\mathrm{h,50}}$. Rather, the goal is to demonstrate the feasibility of estimating $t_{\mathrm{h,50}}$ within an acceptable error range using local galaxy and group properties.  Although the $t_{\mathrm{h,50}}$ serves as a proxy of halo assembly history, it cannot completely capture the entire history of halo assembly. Random merging and stripping events lead to continuous fluctuations in halo assembly history, which may obscure the signals encoded in the present-day properties of galaxies. Moreover, our machine-learning models only employ several observable galaxy properties as input quantities, some of which have non-negligible uncertainties due to measurement, resulting in inevitable deviations in $t_{\mathrm{h,50}}$ estimation results. For example, the $R^2$ score reflected in the test set is not particularly high. In summary, although our model is relatively simplistic and approximate, it effectively categorizes galaxies into groups residing in halos that assembled earlier (higher $t_{\mathrm{h,50}}$) and those in halos that assembled later (lower $t_{\mathrm{h,50}}$).

Unlike halo mass estimation, the relative values of $t_{\mathrm{h,50}}$ that we estimate are more robust than the absolute values. Our derived $t_{\mathrm{h,50}}$ catalog is particularly valuable for comparing galaxies in a relative manner, as it enables investigations into differences among galaxies residing in halos with varying assembly times. Moreover, the uncertainties associated with time resolution in L-GALAXIES are not taken into account in the calculation of $t_{\mathrm{h,50}}$ for the mock data. Therefore, the rank of $t_{\mathrm{h,50}}$ we derived is considerably more reliable than the absolute value, enhancing its utility for relative comparisons among galaxies.

To validate the derived $t_{\mathrm{h,50}}$ for each individual group and the overall $t_{\mathrm{h,50}}$ distributions, we plan to utilize constrained cosmological simulations such as Elucid \citep[e.g.][]{Wang2014,Wang2016,Tweed2017}. Elucid has demonstrated success in reconstructing the formation of large-scale structures and recovering the mass distribution in the local universe as observed by SDSS. It can be effectively employed to trace the assembly history of each halo.

\section{Summary}\label{conclusion}

In addition to the halo mass of the galaxy group, the assembly history of the dark matter halo is also one of the fundamental properties of the halo, which is tightly connected to the properties of the galaxies within. For instance, at the same present-day halo mass, different halo assembly histories can produce significantly different final stellar mass of the central galaxy within. In this work, we explore how to improve the accuracy in estimating halo assembly history (indicated by the halo assembly time $t_{\mathrm{h,50}}$) in observation. We develop machine-learning regression models to estimate $t_{\mathrm{h,50}}$ for low-redshift galaxy groups using only observable quantities in the mock data from the semi-analytic model L-GALAXIES, as it can accurately reproduce the observed stellar mass function in the local universe. Since star-formation quenching will decouple the co-growth of the dark matter and stars, we train our models separately for groups with star-forming central (blue group) and passive central (red group) at each redshift range, and then apply them to each group in the SDSS Yang et al. galaxy group catalog. The main results are summarized below:

(1) The stellar-to-halo mass ratio ($M_*/M_{\mathrm{h}}$) for the centrals has often been used to indicate the halo assembly time $t_{\mathrm{h,50}}$ of the group. Using mock data from the semi-analytic model L-GALAXIES, we find that $M_*/M_{\mathrm{h}}$ shows a significant scatter up to $\sim$ 4 Gyr with $t_{\mathrm{h,50}}$ (Figure \ref{t50SHMR}), with a strong systematic difference between the group with a star-forming central (blue group) and passive central (red group). Hence in addition to $M_*/M_{\mathrm{h}}$, various other group properties are needed to improve the accurate estimation of $t_{\mathrm{h,50}}$.

(2) Using only observable quantities in mock data with added observational uncertainties, we develop machine-learning regression models to estimate $t_{\mathrm{h,50}}$ for galaxy groups within 0.01<$z$<0.20. We train the models separately for stellar-mass-complete and -incomplete samples of blue groups and red groups. We show that our models have successfully recovered the $t_{\mathrm{h,50}}$, with little systematic bias for both blue and red groups. The average uncertainty (RMSE) in $t_{\mathrm{h,50}}$ is $\sim$0.99 Gyr for blue groups, and $\sim$1.32 Gyr for red groups within 0.01<$z$<0.04 (Figure \ref{Regression}). Nevertheless, our models do not perform well for red groups assembled relatively recently (characterized by $t_{\mathrm{h,50}}$ < 5 Gyr, account for less than 5\% of total mock samples), and even including non-observable or hard-to-observe properties, can barely improve the model performance for this small population, probably due to the recent merger of the halos in a random fashion. 

(3) At a given redshift range, the RMSE values for the stellar-mass-incomplete sample models (Figure \ref{Regression004007inc}) are lower than those for the stellar-mass-complete sample models. This trend is likely due to the differing halo mass and stellar mass distributions for stellar-mass-complete and -incomplete samples, as well as mass-dependent RMSE in $t_{\mathrm{h,50}}$ estimation. On average, estimating $t_{\mathrm{h,50}}$ for more massive halos at a given redshift range is generally more challenging (with higher RMSE values), primarily due to more complex assembly histories for these more massive systems (Figure \ref{M-RMSE}). At a given stellar mass, the RMSE for the red group exceeds that of the blue group because the halo mass of the red group is typically larger than that of the blue group \citep{Mandelbaum2006, Zu2016, Luo2018, Bilicki2021}.

(4) Among the input observable quantities to models, we identify the most informative properties that contribute to estimating $t_{\mathrm{h,50}}$ in both blue groups and red groups based on the feature importance rank in the regression models (Figure \ref{FeatureRanking}). Specifically, we find that the stellar age and stellar-to-halo mass ratio of the central galaxy are the most significant features in inferring $t_{\mathrm{h,50}}$. Besides, the total mass of the group and halo mass also play crucial roles.

(5) With careful calibrations of individual observable quantities in the mocks with SDSS observations, we apply these trained regression models to the SDSS Yang et al. groups and accurately derive the $t_{\mathrm{h,50}}$ for each group. To assess the accuracy of the derived $t_{\mathrm{h,50}}$, we show the derived SDSS $t_{\mathrm{h,50}}$ distributions are in reasonable agreement with that in the mocks, particularly for blue groups, which hence strongly supports our methods (Figure \ref{t50eachz}). For SDSS groups, the distributions of $t_{\mathrm{h,50}}$ tend to shift slightly towards lower values as the redshift increases. This is primarily because SDSS groups at lower redshifts contain a substantially higher number of low-mass galaxies, statistically corresponding to low-mass halos. These halos typically assembled earlier and thus exhibit higher $t_{\mathrm{h,50}}$ values than their counterparts at higher redshifts. Additionally, the derived SDSS $t_{\mathrm{h,50}}$ distributions at higher redshifts show progressively larger deviations from that in the mocks, due to the increasingly larger observational uncertainties in group findings and property estimations.

The halo assembly history, together with the halo masses, makes an important step forward in studying the halo-galaxy connections. Future larger and deeper sky surveys will make it possible to incorporate additional group and galaxy properties in our method to further improve the accuracy in the measurements of both halo mass and halo assembly history of individual galaxy groups. These properties may include the size and geometry of the groups, the density profile of the groups \citep[e.g.,][]{Newman2015, Wang2024}, the kinematics of the centrals \citep[e.g.,][]{Cappellari2016, Peng2020, Renzini2020}, etc.

\begin{acknowledgments}
We thank the anonymous referee for useful comments that have improved the paper. Y.J.P. and C.Q.L. acknowledge the support from the National Key R\&D Program of China Grant 2022YFF0503401, National Science Foundation of China (NSFC) grant Nos. 12125301, 12192220, 12192222, and the science research grants from the China Manned Space Project with No. CMS-CSST-2021-A07. E.W. acknowledges the support of National Science Foundation of China and the Start-up Fund of the University of Science and Technology of China, Grant No. GG2030007025 and Grant No. KY2030000200. Q.S.G. acknowledges the support from the National Science Foundation of China (NSFC) grant Nos. 12192222, 12192220 and 12121003. J.D. acknowledges the support of National Science Foundation of China (NSFC) grant Nos. 12303010. This work is extensively supported by the High-performance Computing Platform of Peking University, China.
\end{acknowledgments}

\section*{Data Availability}
All derived data in this work will be fully publicly available in our following data release paper, together with various other derived properties of the halos and galaxies.


\appendix

\section{Model performance for other redshift range}\label{app}
\renewcommand{\appendixname}{Appendix~\Alph{section}}
\setcounter{figure}{0} 
\renewcommand{\thefigure}{A\arabic{figure}}

\begin{figure*}[htb] 
	\centering
	\includegraphics[width=0.48\textwidth]{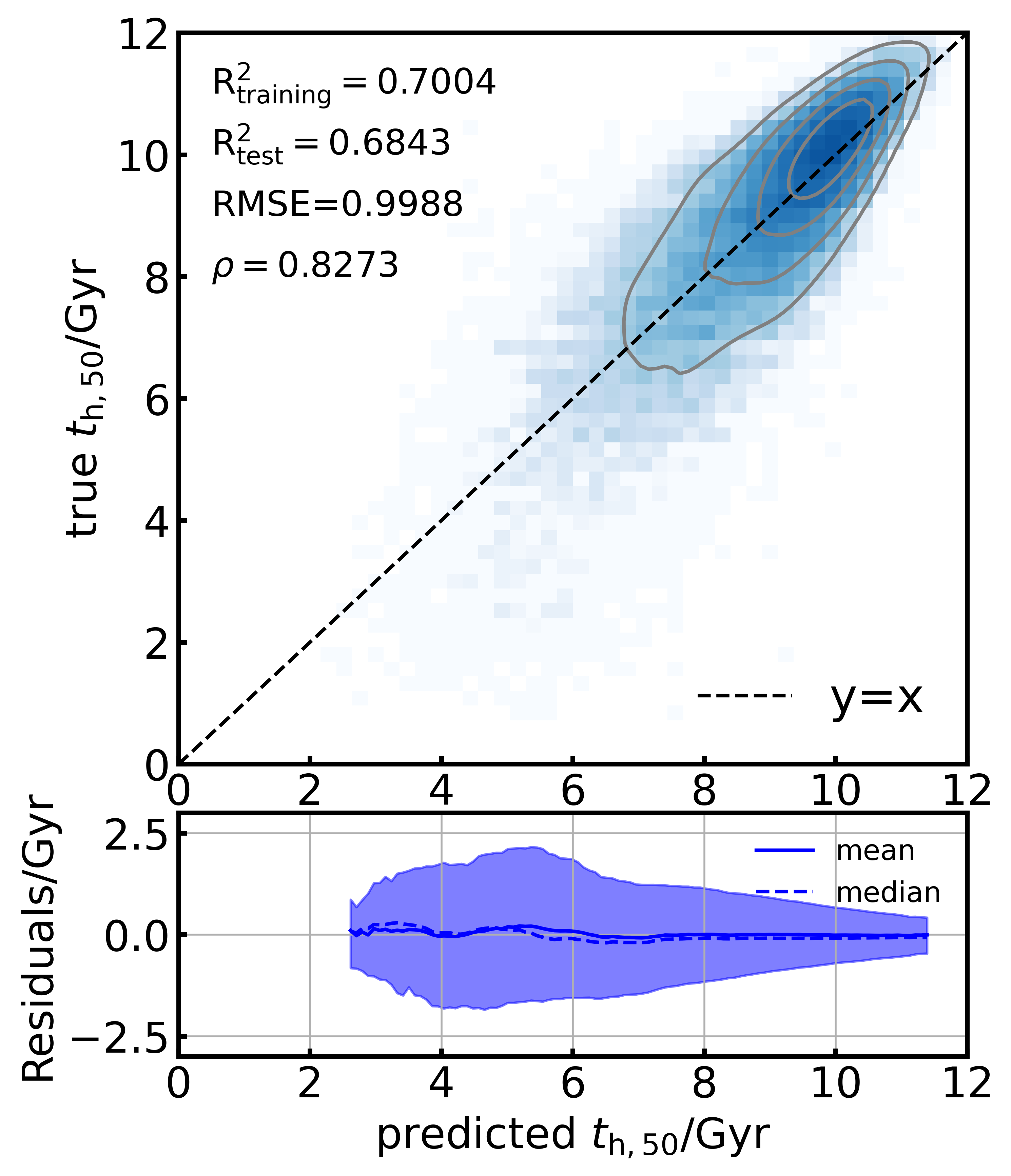}
	\includegraphics[width=0.48\textwidth]{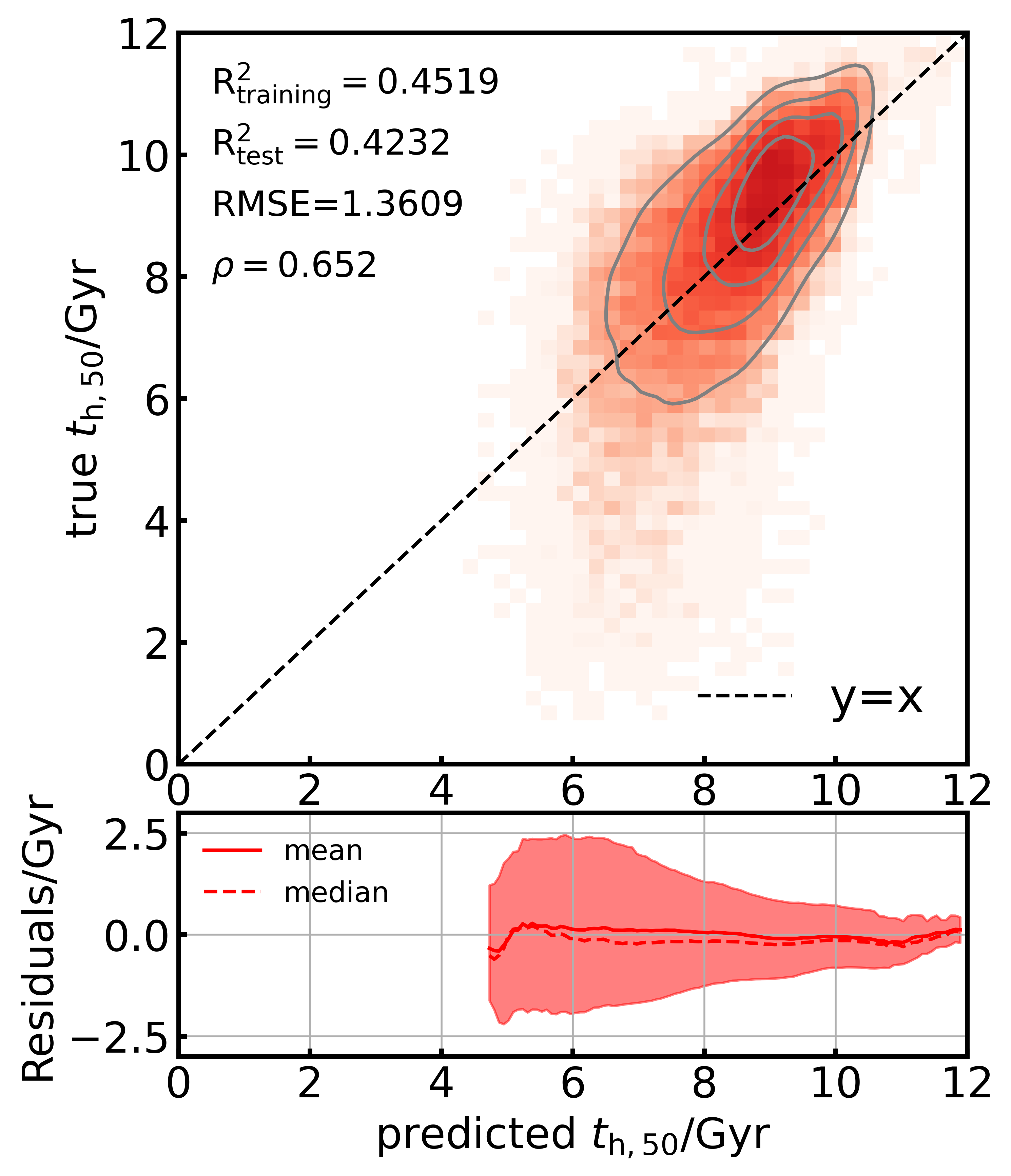}
	\caption{Similar to Figure \ref{Regression}, but for the test sample of the stellar-mass-complete sample models within 0.04<$z$<0.07.}
	\label{Regression004007com}
\end{figure*}

Figure \ref{Regression004007com} illustrates the performance of regression models on test samples for the stellar-mass-complete sample within 0.04<$z$<0.07. The RMSE values are $\sim$ 0.9988(1.3609) Gyr for blue(red) group samples. The sample selection in the mock data is similar to that in SDSS, as shown in Figure \ref{MassRedshift}. First, compared with models for stellar-mass-complete sample within $0.01<z<0.04$, as shown in Figure \ref{Regression}, the RMSE values are slightly higher at higher redshift, primarily due to the presence of more higher-mass galaxies (see mass-dependent RMSE in Figure \ref{M-RMSE}). Second, compared with models for the stellar-mass-incomplete sample at the same redshift range, as shown in Figure \ref{Regression004007inc}, the RMSE values are higher for the stellar-mass-complete sample, primarily due to higher-mass galaxies (see mass-dependent RMSE in Figure \ref{M-RMSE}). Third, the general trend of RMSE across different regression models is higher for stellar-mass-complete samples than for stellar-mass-incomplete samples and higher for samples at higher redshifts than for those in lower redshifts, as a result of the stellar mass distributions in Figure \ref{MassRedshift}.

\bibliography{T50_ref}{}   
\bibliographystyle{aasjournal}

\end{document}